\documentclass{IEEEtran}
\usepackage{cite}
\usepackage{amsmath,amssymb,amsfonts}
\usepackage{graphicx}
\usepackage{textcomp,nicefrac}
\usepackage{subfig}
\usepackage{mathtools}
\usepackage{booktabs}
\usepackage{tabularx}

\graphicspath{{figures/}{}} %Setting the graphicspath
\def\BibTeX{{\rm B\kern-.05em{\sc i\kern-.025em b}\kern-.08em
T\kern-.1667em\lower.7ex\hbox{E}\kern-.125emX}}
\begin{document}
\title{Design, fabrication and preliminary characterization of a novel 3D-Trench Sensor implemented in 8-inch CMOS-Compatible Technology}
\author{Manwen Liu, Huimin Ji, Wenzheng Cheng, Chuan Liao, Le Zhang, Zheng Li, Bo Tang, Peng Zhang, Wenjuan Xiong, Trevor Vickey, E. Giulio Villani, Zhihua Li, Dengfeng Zhang, and Jun Luo
\thanks{
This paper is submitted on \today.
This work is supported by the National Key R\&D Program of China under Grant 2023YFF0719600, General Program of National Natural Science Foundation of China under Grant 12375188, and the  Science  and  Technology  Facilities  Council  (STFC)  of  United  Kingdom  under  grants ST/S000747/1 and ST/W000547/1.
We are also grateful for the support with equipment and technical personnel by the IMECAS, the  School  of  Mathematical  and  Physical  Sciences of the University of Sheffield,  and the Rutherford Appleton Laboratory.
(Corresponding author: Manwen Liu, Chuan Liao, Zhihua Li and Dengfeng Zhang.)}
\thanks{Manwen Liu and Zhihua Li are with the Key Laboratory of Fabrication Technologies for Integrated Circuits of Chinese Academy of Sciences and the Institute of Microelectronics of Chinese Academy of Sciences, Beijing 100029, China (e-mail: liumanwen@ime.ac.cn, lizhihua@ime.ac.cn).}
\thanks{Chuan Liao is with the International Center for Quantum-field Measurement Systems for Studies of the Universe and Particles(QUP), High Energy Accelerator Research Organization(KEK), Japan (e-mail: liao@post.kek.jp).}
\thanks{Dengfeng Zhang is with the School of Mathematical and Physical Sciences, The University of Sheffield, Sheffield, S10 2TN, UK (e-mail: dengfeng.zhang@cern.ch).}
\thanks{Jun Luo, Bo Tang, Wenjuan Xiong and Peng Zhang are with the Key Laboratory of Fabrication Technologies for Integrated Circuits of the Chinese Academy of Sciences and the Institute of Microelectronics of the Chinese Academy of Sciences (IMECAS) in Beijing, China.}
\thanks{Huimin Ji, Le Zhang and Wenzheng Cheng are with the Key Laboratory of Fabrication Technologies for Integrated Circuits of the Chinese Academy of Sciences, Institute of Microelectronics of the Chinese Academy of Sciences (IMECAS) and the University of Chinese Academy of Sciences, Beijing, China.}
\thanks{Zheng Li is with Ludong University in Yantai and the Institute of Microelectronics of the Chinese Academy of Sciences (IMECAS), Beijing, China.}
\thanks{Trevor Vickey is with the School of Mathematical and Physical Sciences, The University of Sheffield, Sheffield, S10 2TN, UK.}
\thanks{E. Giulio Villani is with the Particle Physics Department, Rutherford Appleton Laboratory of the Science and Technology Facilities Council (STFC), Oxfordshire, UK.}
}

\maketitle

\begin{abstract}
The 3D silicon sensor has demonstrated excellent performances (signal collection, detection efficiency, power consumption, etc.) comparable or even better with respect to the traditional planar sensor, especially after the high irradiation fluence, mainly due to the shorter drift length of the generated carriers.
These characteristics have made it the most attractive technology for the detection and track reconstruction of charged particles for the High Energy Physics (HEP).
In addition, its application is also being explored in astronomy, microdosimetry and medical imaging.
This paper presents the design and fabrication of a novel 3D-Trench sensor which features an enclosed deep trench surrounding the central columnar cathode.
This novel sensor has been fabricated on the 8 inch COMS pilot line at the Institute of Microelectronics of the Chinese Academy of Sciences (IMECAS) where ultra-narrow etch width of 0.5 $\mu$m and the ultra-high depth-to-width ratio (aspect ratio) (\textgreater 70) have been achieved.
Its preliminary simulation and characterization results including electrostatic potential, electric field, Current-Voltage (IV), Capacitance-Voltage (CV), Charge Collection Efficiency (CCE) and Timing Performance before irradiation will be presented in this paper.
\end{abstract}


%---------------
\section{Introduction}
\label{sec:introduction}

\IEEEPARstart{S}{ince} its proposal by S. Parker et al. in 1998~\cite{PARKER1997328, 785737}, the 3D sensor has undergone extensive research and has found diverse applications in high-energy physics experiments (the High-Luminosity LHC)~\cite{Heggelund_2022, Furelos_2017, Lange_2016}, microdosimetry for proton and heavy-ion treatment~\cite{10.1063/1.4926962, 9339896, s41598-022-14940-1}, X-ray spectroscopy in computed tomography (CT) scanners~\cite{10238431}, and astroparticle detection in deep space radiation environments~\cite{8089762}. These applications benefit from the sensor's outstanding features, including short collection distances, fast collection times, low depletion voltages, radiation hardness, and the ability to provide position resolution of the order of microns.
Furthermore, SiO\textsubscript{2} 3D trench-isolation technology has been employed in avalanche diodes to achieve small pixel pitch and high fill-factor, enhancing their performance in imaging and particle detection~\cite{9081916, 10019414}.

In the past two decades, institutes such as IMB-CNM~\cite{PELLEGRINI201969, Terzo_2022, DIEHL2024169517}, SINTEF~\cite{Heggelund_2022, Dorholt_2018, Terzo_2021, 6154334}, University of Trento~\cite{chips2020006, Terzo_2021}, INFN~\cite{GRENIER201133}, FBK~\cite{6522814, 4696595, 5873785, Sultan_2017}, BNL~\cite{LI201190}, and others have conducted extensive research on 3D sensors with varying electrode thicknesses, geometries, aimed at wide ranges of scientific applications.
The aforementioned research has provided valuable design insights and techniques for the fabrication process for the development of 3D sensors aimed at different applications. Recent investigation has shown excellent timing resolution of less than 100 ps~\cite{Diehl_2022}.

The fabrication process of 3D sensors presents several challenges, such as the precise shape of the electrodes, their size and depth-to-width ratio (aspect ratio) of the electrodes.
For example, the ATLAS Phase-II upgrade for the HL-LHC~\cite{Lange_2016} introduces additional complexities, with the new ITk (Inner Tracker) module necessitating a reduction in pixel unit size and substrate thickness~\cite{TERZO2020164587, DALLABETTA2016388}.
Furthermore, the production of wafers with thicknesses below 100 micrometers presents significant fabrication challenges, including wafer bending and alignment, posing new difficulties to the manufacturing process.

In this work, an innovative 3D-Trench sensor was designed and fabricated on the epitaxial layer (EPI) grown on 8-inch wafer. This design maintains a thin effective volume while mitigating the fabrication risks associated with thin silicon substrates. The sensor was fabricated using the 8-inch CMOS process platform at the Institute of Microelectronics of the Chinese Academy of Sciences (IMECAS). Notably, we have achieved an aspect ratio of 70:1 with an electrode width of 0.5 $\mu$m, significantly enhancing the sensor's fill factor.
The manufacture technology based on the 8-inch CMOS compatible process lays a solid foundation for future large-scale production and application due to its scalability, flexibility, cost-effectiveness, maturity, and compatibility.
Additionally, the electrical isolation of the 3D trench electrode simplifies the guard ring design compared to traditional 2D or 3D sensors.
The designed 3D-Trench sensor exhibits low depletion voltage and fast signal collection.
Our theoretical analysis is grounded in the frameworks of the 3D-Trench electrode sensor~\cite{LI201190} and near-hemispherical electrode sensor~\cite{Liu_2021}, providing a robust foundation for our findings.

This paper is organized as follows.
Section \ref{sec:design} gives a description of the design of the silicon 3D-Trench sensor.
In section \ref{sec:fabrication}, a description of the fabrication process of the devices will be given.
Section \ref{sec:result} will present test and simulation results, including Current-Voltage (IV), Capacitance-Voltage (CV), Charge Collection Efficiency (CCE), Time Resolution and Rise Time.
Finally, in section \ref{sec:conclusion}, preliminary conclusions and a description of the next steps for this project will be given.

%---------------
\section{Design of the 3D-Trench Sensor}
\label{sec:design}
\begin{figure}[htbp]
\centering
\includegraphics[width=\columnwidth]{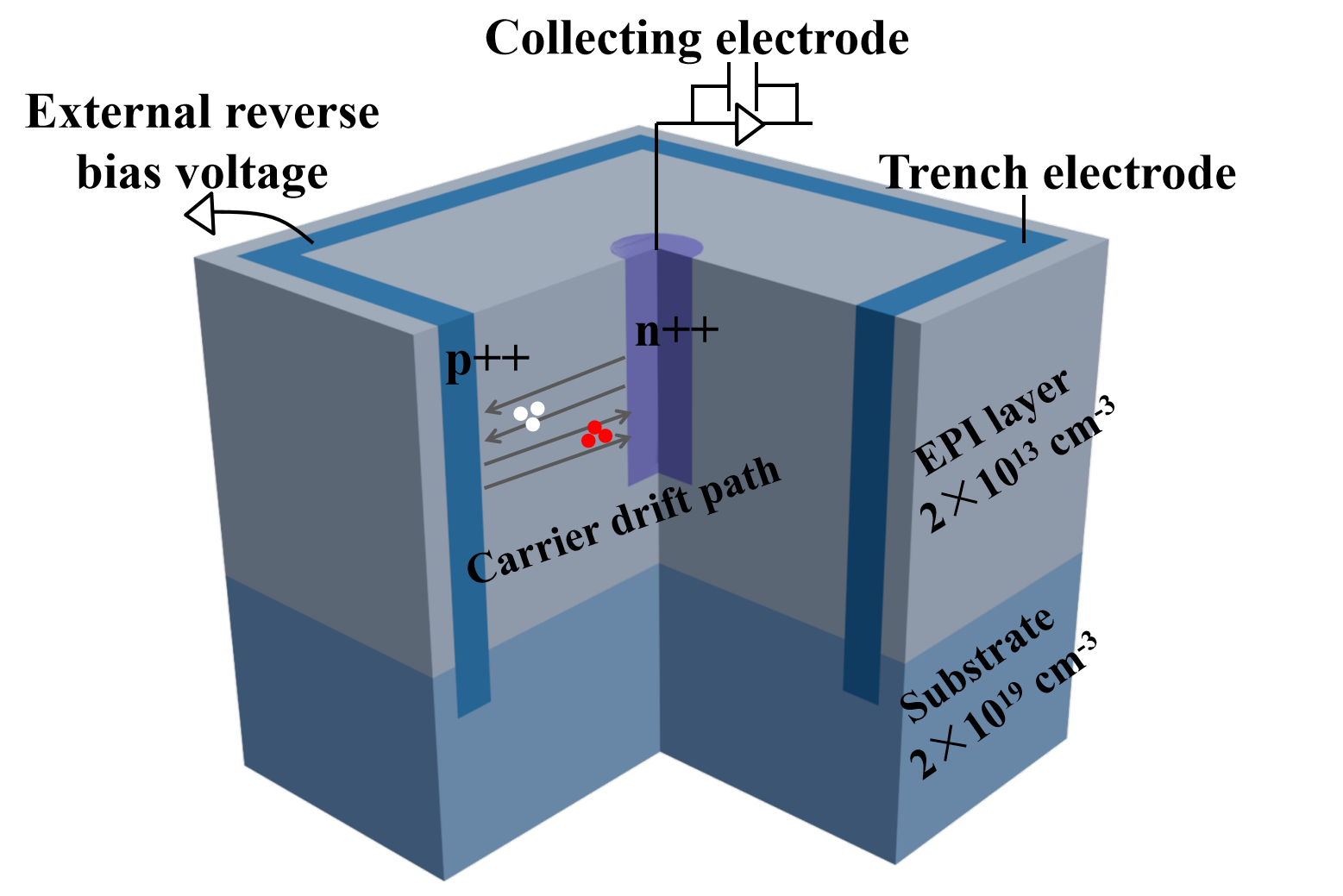}
\caption{The schematic of the 3D-Trench Sensor.}
\label{fig:schematic}
\end{figure}

In the novel design of the 3D-Trench sensor, the columnar cathode is surrounded by a deeply etched trench, instead of columnar anodes, as in traditional 3D pixel sensor~\cite{PARKER1997328}. The shape of the trench can be square or circular.
Fig. \ref{fig:schematic} shows the schematic of the 3D-Trench sensor.
The 3D-Trench sensor devices are fabricated on 8-inch silicon wafer, the sensitive layer is a high resistivity \textit{p}-type epitaxial layer with a thickness of 30 $\mu$m, grown on low-resistivity \textit{p}-type substrate.
Both the enclosed 3D-Trench anode and the central columnar cathode are achieved by etching the epitaxial layer from the front side. The 3D-Trench anode will fully penetrate the epitaxial layer and into the substrate by $\sim$5 $\mu$m, but the etching for the central columnar cathode will stop at 5 to 10 $\mu$m from the substrate. More details on the fabrication are presented in Section \ref{sec:fabrication}.
Both columnar and trench electrodes have a width of 0.5-2 $\mu$m with an achieved aspect ratio of 70:1 or larger, which suppresses the dead areas significantly.
In this design, the \textit{pn} junction is formed near the central electrode and the depletion region develops from the central cathode up to the trench wall.

\begin{figure}[htbp]
\centering
\subfloat[]{\includegraphics[width=0.5\columnwidth]{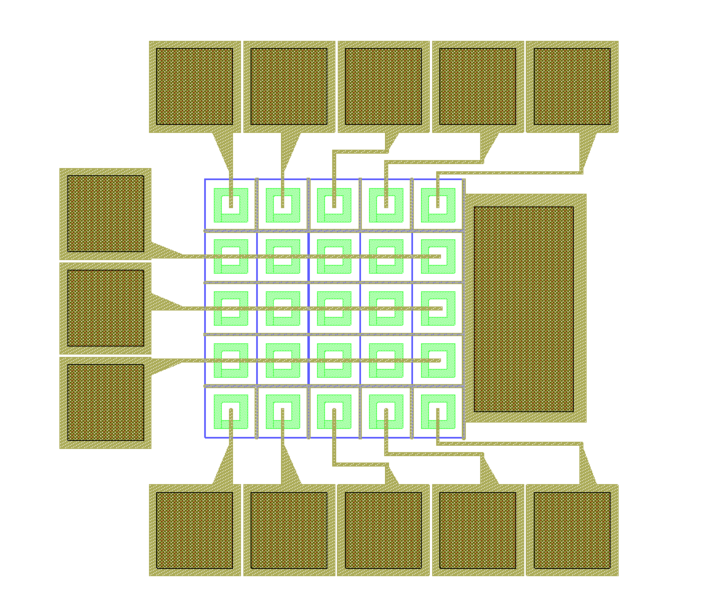} \label{fig:layout_5by5_pixel_array}}
\subfloat[]{\includegraphics[width=0.5\columnwidth]{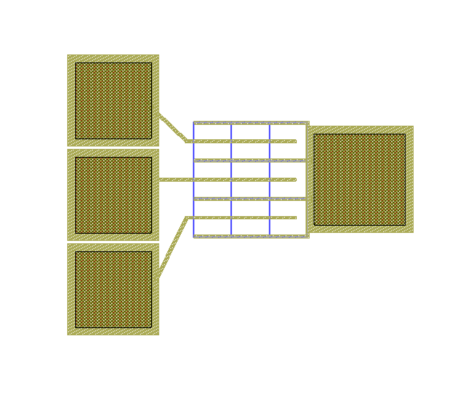} \label{fig:layout_3by3_pixel_array}}
\caption{Layouts of the 3D-Trench pixel array, (a) A21-21: 5$\times$5 pixel array with the pixel size of 35$\times$35 $\mu$m, (b) A21-43: 3$\times$3 pixel array with the pixel size of 25$\times$25 $\mu$m.}
\label{fig:device_layout}
\end{figure}

We have designed different device layouts differing in the shape of the 3D trench and pixel size.
Fig. \ref{fig:device_layout} shows two example layouts: A21-21 (5$\times$5 pixel array with the pixel size of 35$\times$35 $\mu$m) and A21-43 (3$\times$3 pixel array with the pixel size of 25$\times$25 $\mu$m). Their details are listed in Table \ref{table:device_layout}.

\begin{table}[htbp]
\centering
\caption{Layouts of 3D-Trench pixel devices: A21-21 and A21-43.}
\label{table:device_layout}
%\begin{tabular}{ccccc}
\begin{tabularx}{0.5\textwidth}{>{\centering\arraybackslash}X >{\centering\arraybackslash}X >{\centering\arraybackslash}X >{\centering\arraybackslash}X >{\centering\arraybackslash}X}
\toprule
%Layout & Array & Trench Shape & Pixel Size [$\mu$m] & Electrode Width [$\mu$m]\\
Layout & Trench Shape & Electrode Width [$\mu$m] & Array & Pixel Size [$\mu$m] \\
\midrule
A21-21 & Square & 0.5 & 5$\times$5 & 35$\times$35 \\
A21-43 & Square & 0.5 & 3$\times$3 & 25$\times$25 \\
\bottomrule
\end{tabularx}
\end{table}

%---------------
\section{Fabrication of the 3D-Trench Sensor}
\label{sec:fabrication}
The 3D-Trench sensors were fabricated on 8$''$ EPI wafer based on the 0.5 $\mu$m CMOS technology at IMECAS in Beijing.
The epitaxial layer is \textit{p}-type silicon with a doping concentration of $2\times10^{13}$ cm\textsuperscript{-3} and has a thickness of 30 $\mu$m.
The substrate is \textit{p}-type silicon with a thickness of 695 $\mu$m and a doping concentration of $2\times10^{19}$ cm\textsuperscript{-3}.

\begin{figure}[htbp]
\centering
\includegraphics[width=\columnwidth]{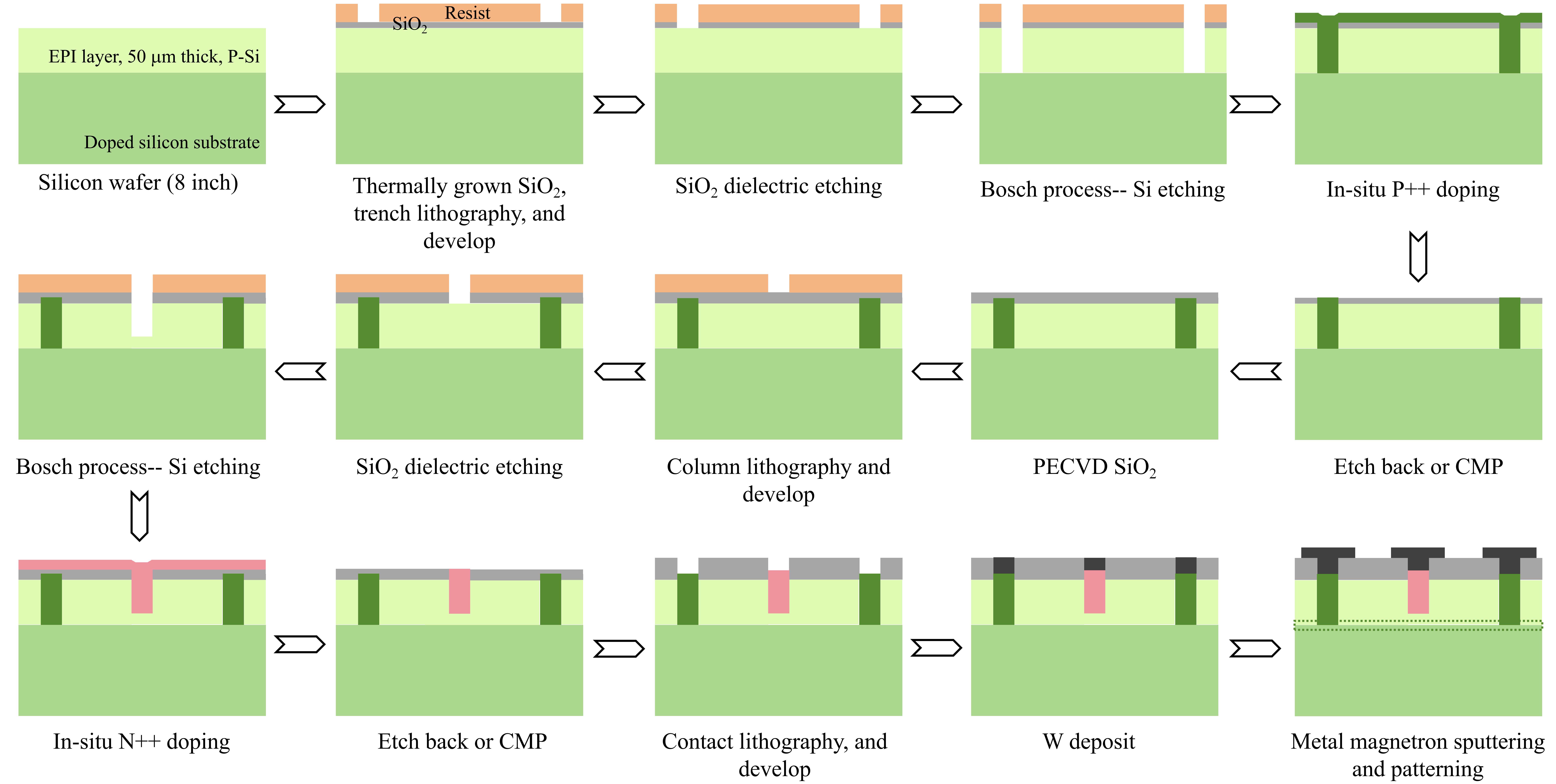}
\caption{The fabrication process flow of the 3D-Trench Sensor at IMECAS.}
\label{fig:fabricationflow}
\end{figure}

The full fabrication process flow is shown in Fig. \ref{fig:fabricationflow}.
The fabrication started with the thermal growth of 700 nm oxide layer on the wafer. This oxide layer provides surface passivation and serves as a hard mask for the subsequent silicon deep etch process.
Next, a layer of photoresist was spin-coated onto the wafer surface, followed by the photolithography and etching procedures to open a 0.5-2 $\mu$m width window through the oxide for the trench etching of the next step.
Next, the Bosch Deep Reactive Ion Etching (DRIE) procedure was performed in the windows opened in the previous step to achieve a 35 $\mu$m deep trench.
After the trench  etching, the in-situ boron diffusion was performed in the etched trench to form a \textit{p}-type enclosed 3D anode and filled the trench with conductive polysilicon.
Thereafter, the polysilicon on the surface was removed and the 3D-Trench anode was obtained.
Before etching the central pillar for the cathode, another oxide layer was deposited on the surface using the Plasma-Enhanced Chemical Vapor Deposition (PECVD) technique.
The same procedures were then performed to form the central columnar cathode, except that the etching depth of the central pillar was shallower (25 $\mu$m) and the phosphorus diffusion was performed to form a \textit{pn} junction.
The last steps consisted of stripping the photoresist on the surface and repeat the photolithography for the electrode metal contact formation.
were to strip the photoresist on the surface and repeat the photolithography for electrode metal contact.
The electrode contact was formed by the tungsten (W) deposit and magnetron sputtering of Al.

\begin{figure}[htbp]
\centering
\subfloat[]{\includegraphics[width=0.5\columnwidth]{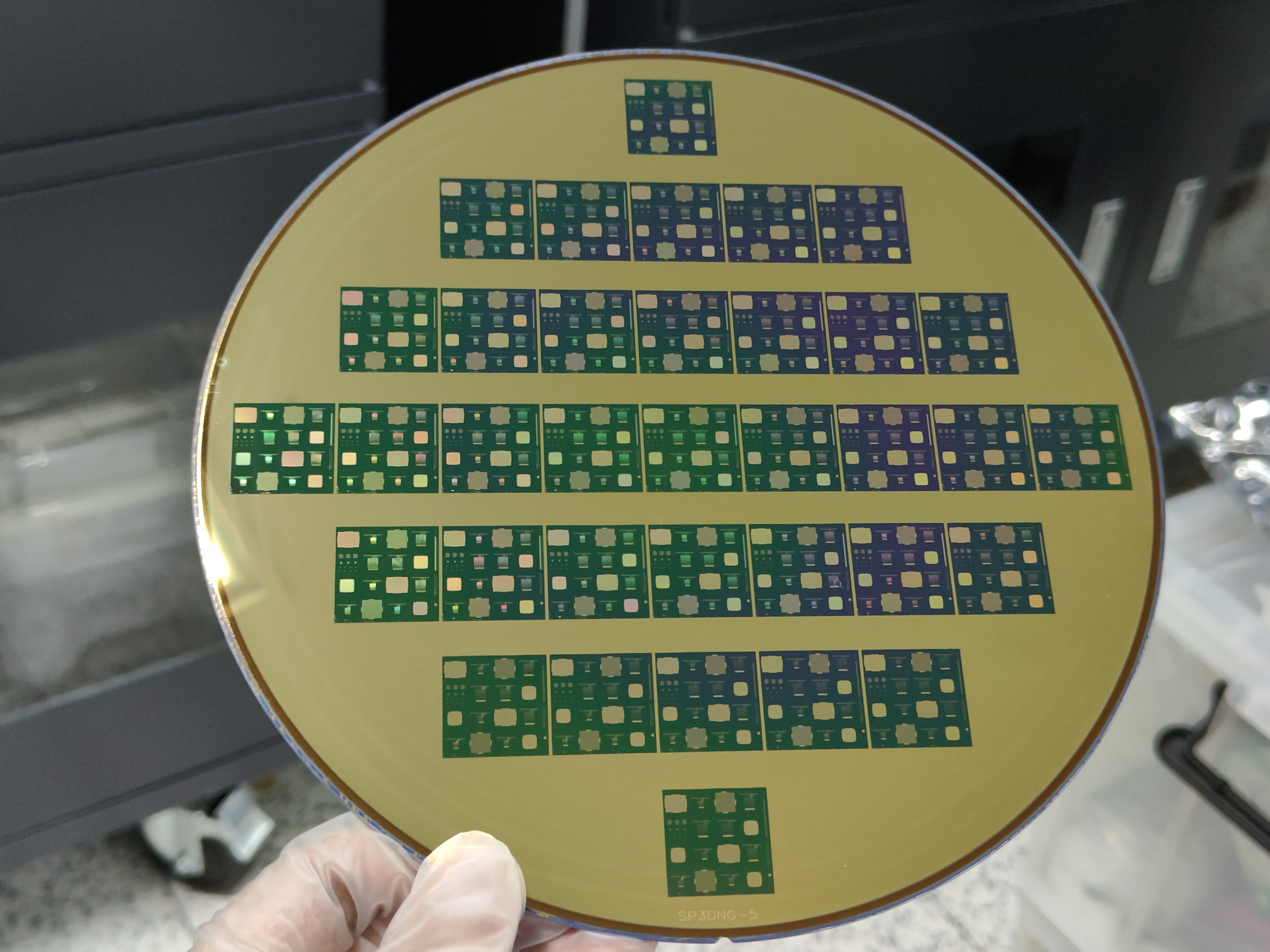} \label{fig:wafer}}
\subfloat[]{\includegraphics[width=0.5\columnwidth]{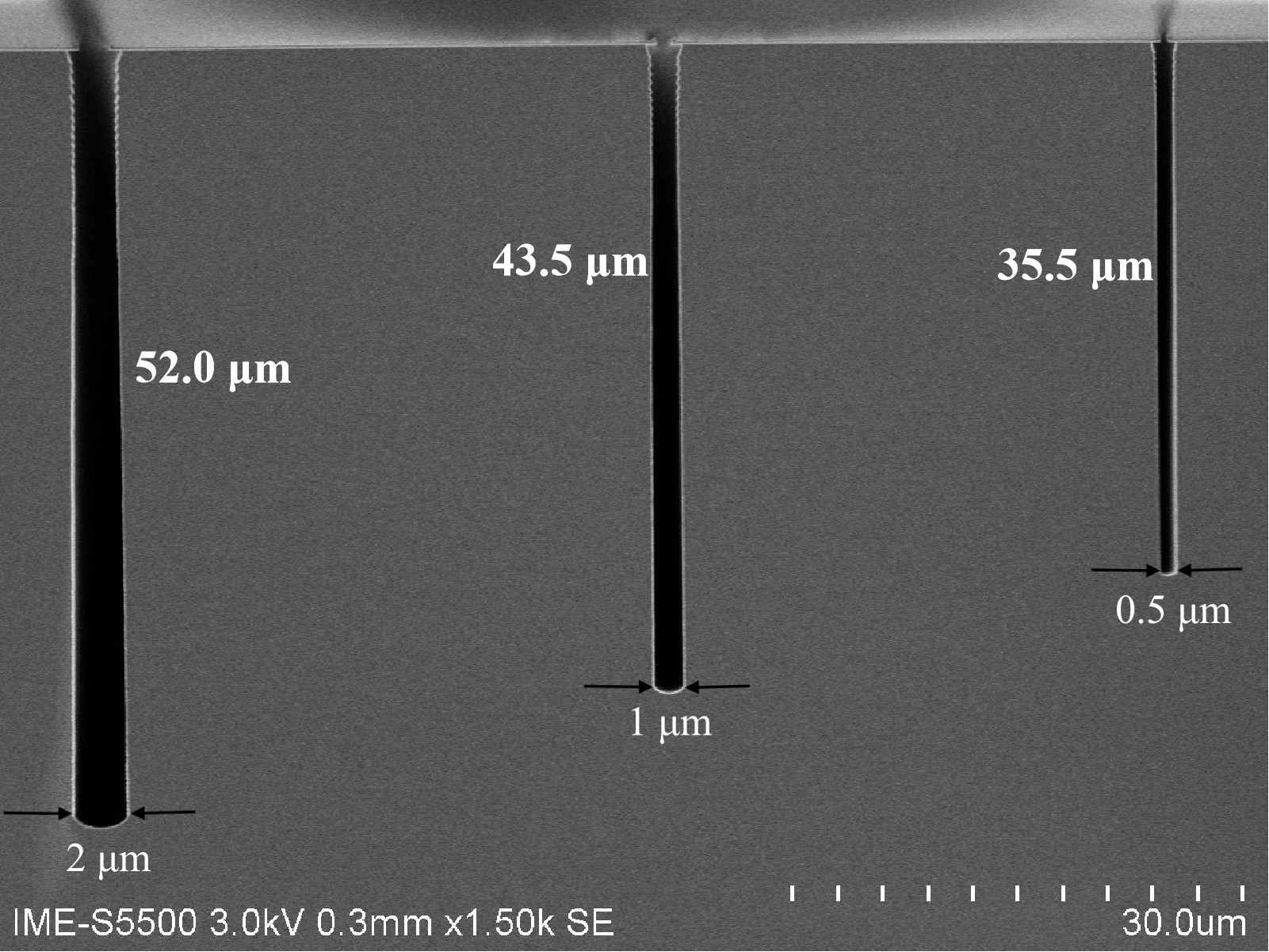} \label{fig:sem_electrode}}
\\
\subfloat[]{\includegraphics[width=0.5\columnwidth]{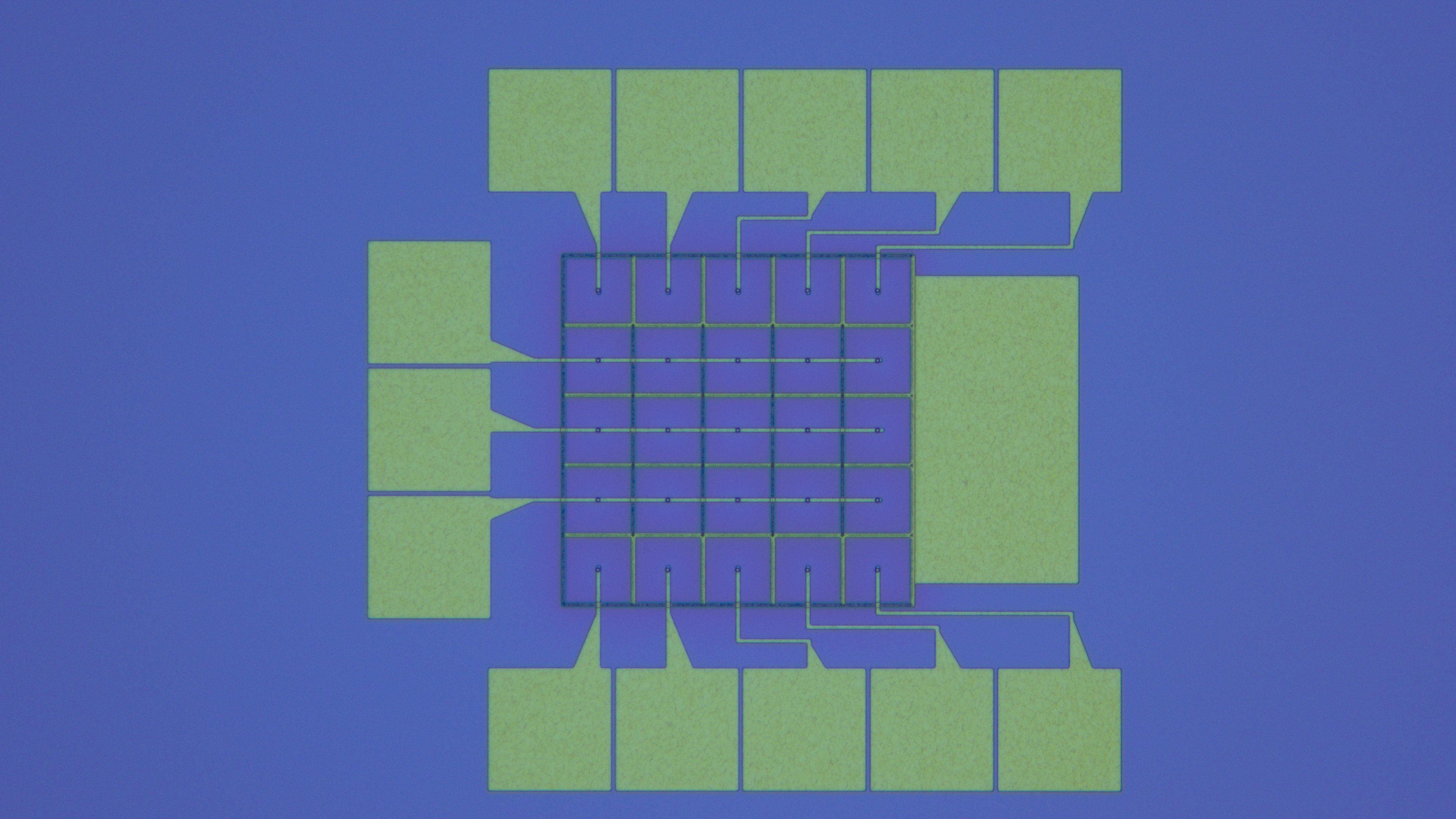} \label{fig:sem_5by5_pixel_array}}
\subfloat[]{\includegraphics[width=0.5\columnwidth]{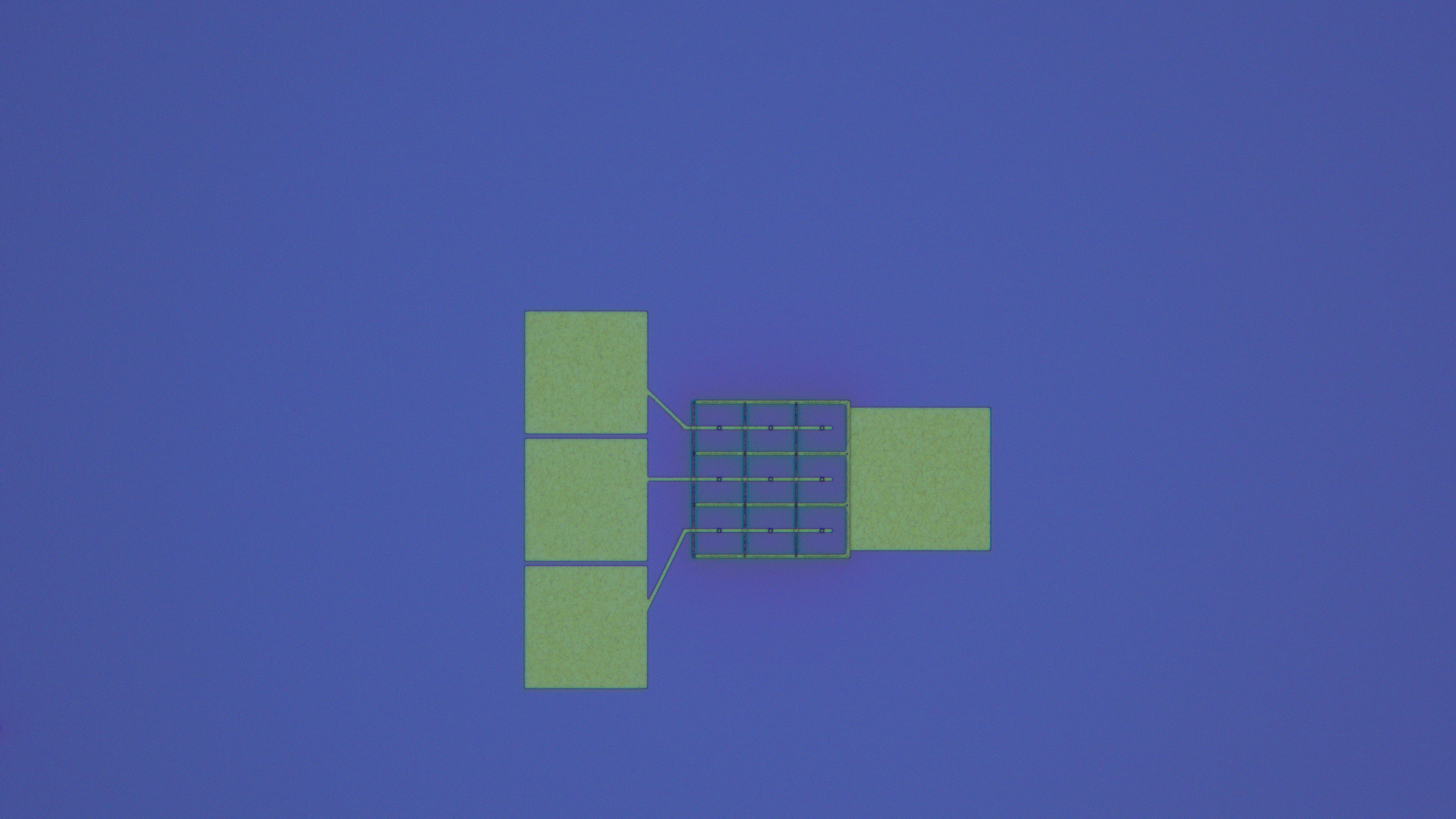} \label{fig:sem_3by3_pixel_array}}
\caption{(a) The final fabricated 8$''$ wafer at IMECAS, (b) the SEM micrograph in cut view of the 3D trench, (c) the microphotographs of the surface of A21-21: 5$\times$5 pixel array with pixel size of 35$\times$35 $\mu$m and (d) A21-43: 3$\times$3 pixel array with pixel size of 25$\times$25 $\mu$m.}
\end{figure}

The final fabricated wafer is shown in Fig. \ref{fig:wafer}.
Fig. \ref{fig:sem_electrode} shows the Scanning Electron Microscopy (SEM) micrography of the etched trench, showing the achieved minimum trench width of $\sim$0.5 $\mu$m and a maximum aspect ratio exceeding 70.
Fig. \ref{fig:sem_5by5_pixel_array} and \ref{fig:sem_3by3_pixel_array} show the microphotographs of the 5$\times$5 pixel array with pixel size of 35$\times$35 $\mu$m and 3$\times$3 pixel array with pixel size of 25$\times$25 $\mu$m, respectively.

%---------------
\section{Characterization Measurement and Simulation of the Device}
\label{sec:result}

\begin{figure}[htbp]
\centering
\subfloat[]{\includegraphics[width=0.5\columnwidth]{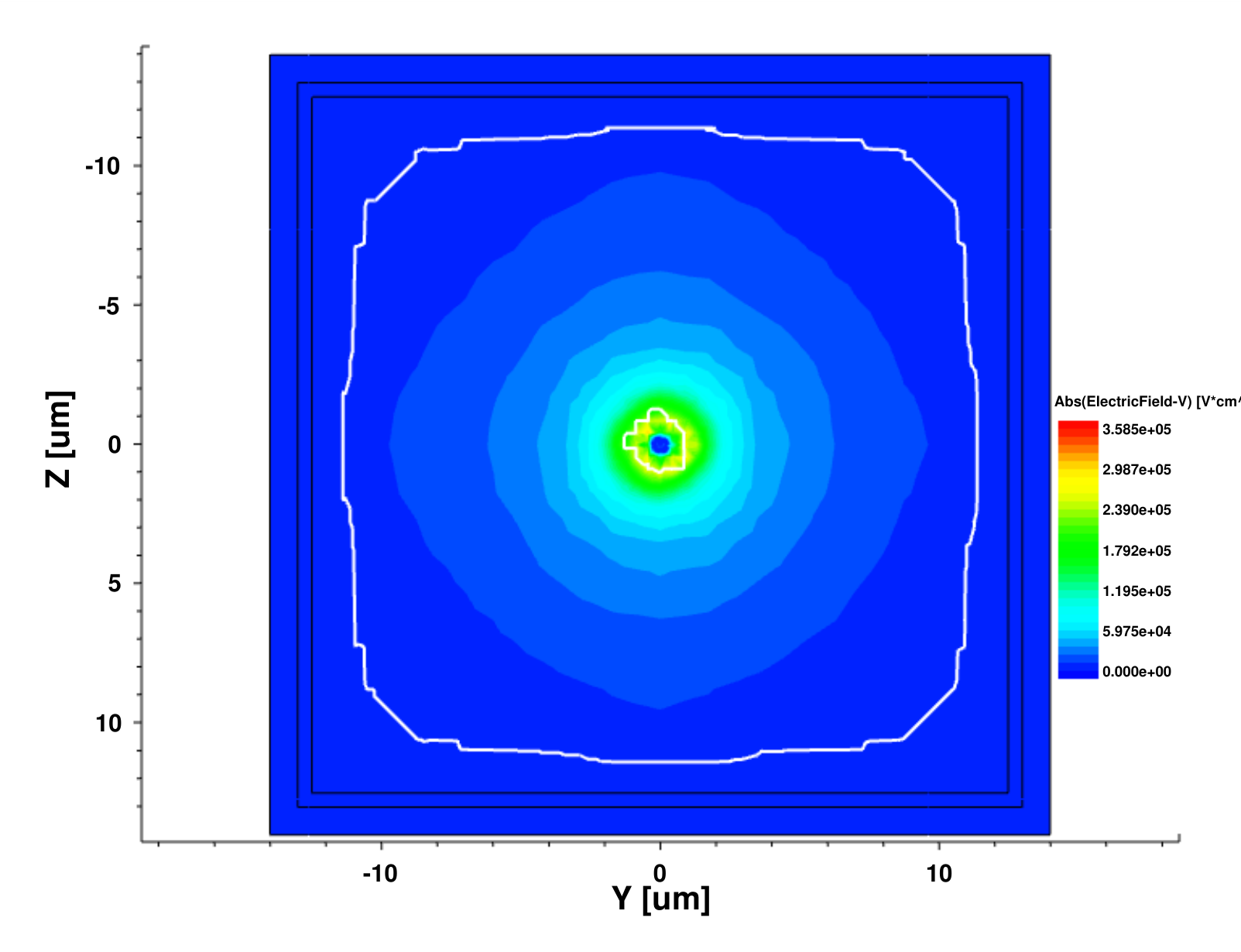} \label{fig:electricfield_a43}}
\subfloat[]{\includegraphics[width=0.5\columnwidth]{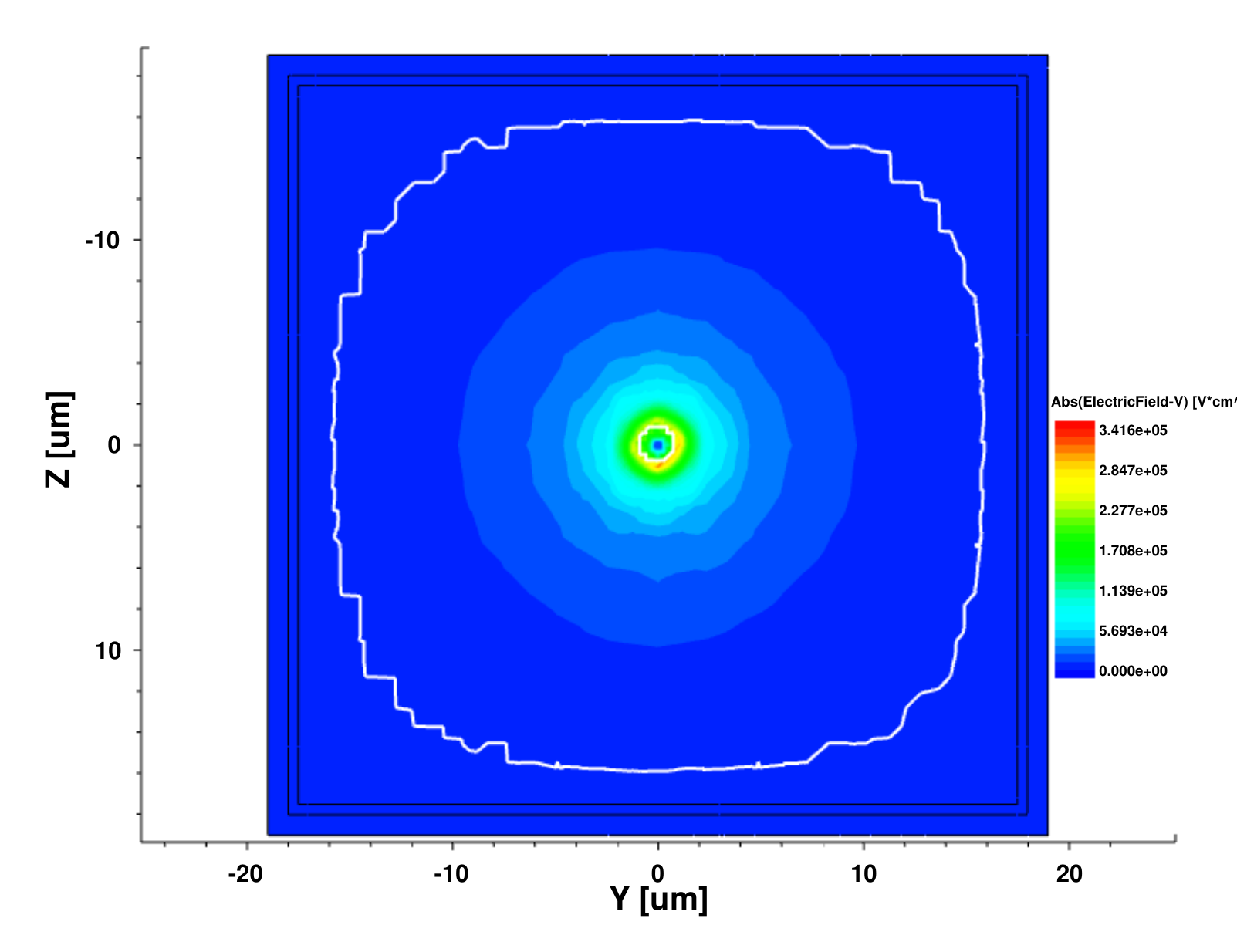} \label{fig:electricpotential_a21}}
\caption{The simulated distributions of the electric field of (a) A21-43 and (b) A21-21 at the reverse bias voltage of 50 V.}
\label{fig:electricfield}
\end{figure}

\begin{figure}[htbp]
\centering
\subfloat[]{\includegraphics[width=0.5\columnwidth]{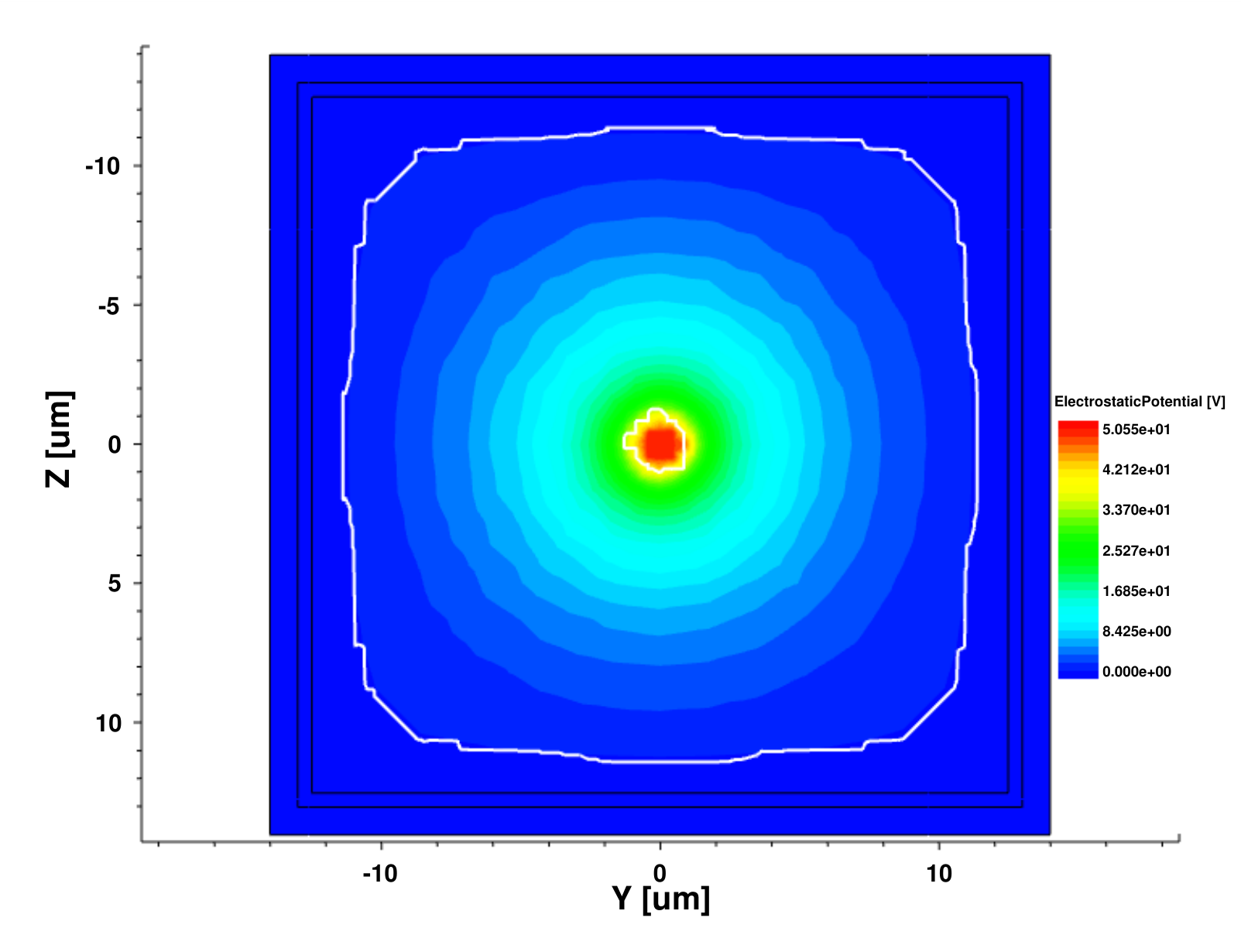} \label{fig:electricpotential_a43}}
\subfloat[]{\includegraphics[width=0.5\columnwidth]{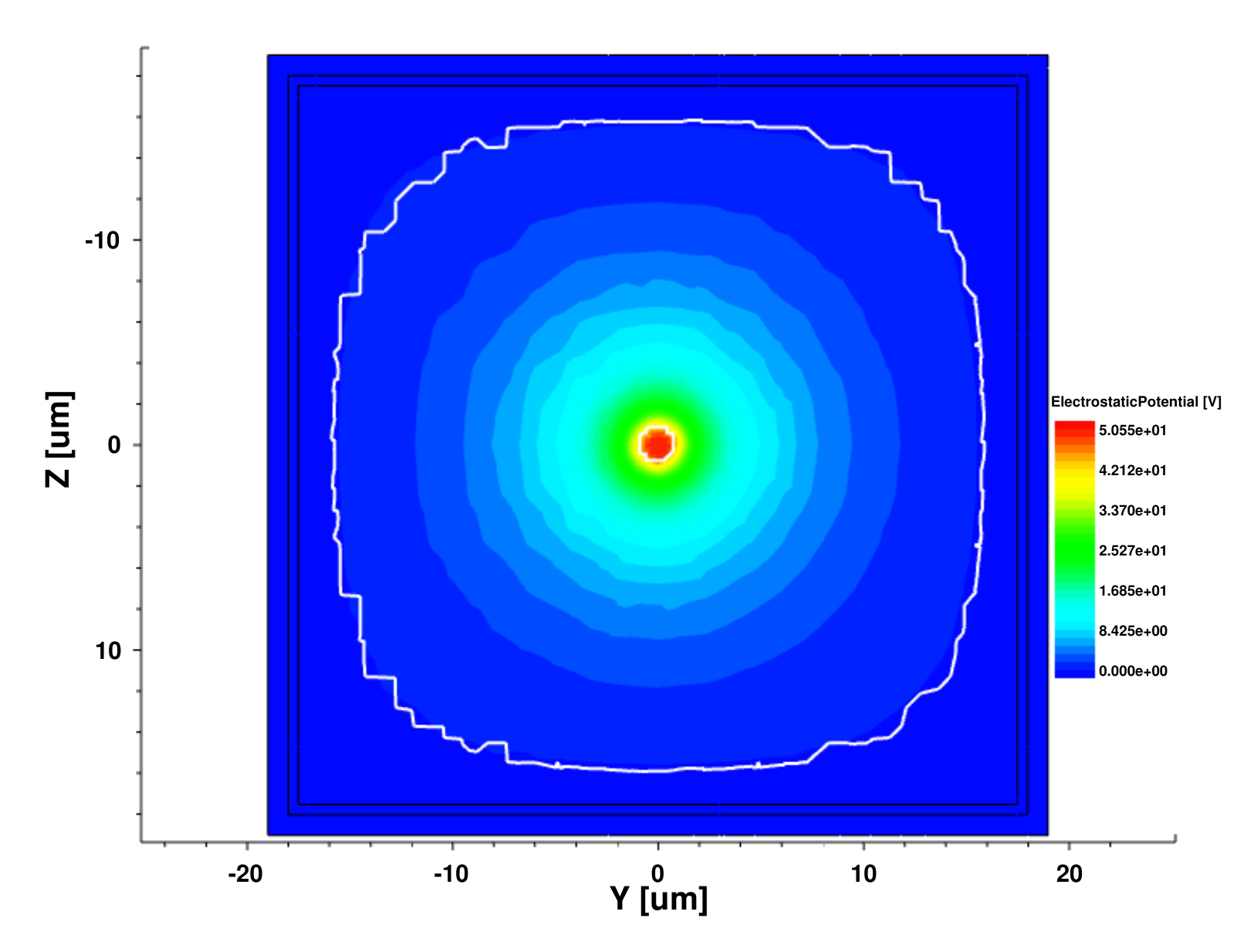} \label{fig:electricpotential_a21}}
\caption{The simulated distributions of the electrostatic potential of (a) A21-43 and (b) A21-21 at the reverse bias voltage of 40 V.}
\label{fig:electricpotential}
\end{figure}

\subsection{Electrical characteristics of the 3D-Trench sensor}
\label{subsection:iv_cv}

The 3D Simulation was performed using the Synopsys TCAD~\cite{Synopsys}.
%In our simulation, a 3D simulated structure in TCAD was obtained from a 2D one by performing a rotation along the x vertical axis.
Fig. \ref{fig:electricfield} and \ref{fig:electricpotential} show the simulated distributions of the electric field and electrostatic potential at the bottom of the cathode of the 3D-Trench Sensor with the pixel size of 25 and 35 $\mu$m at the reverse bias voltage of 50 V, a voltage sufficient to guarantee full depletion.
The maximum electric field is near the bottom of the cathode and the maximum values are $\sim$3.5$\times 10^{5}$ V/cm at the bias voltage of 50 V, high enough to ignite impact ionisation.

The Current-Voltage (IV) and Capacitance-Voltage (CV) characteristics in the reverse biasing mode of the 3D-Trench pixel devices were measured at the Institute of Microelectronics of the Chinese Academy of Sciences (IMECAS) at room temperature (25 $^{\circ}$C) in a clean room.
The setup for the measurement of IV and CV characteristics consists of a Keysight B1505A Power Device Analyzer coupled to a probe station and Keysight B2902A Source/Measure Unit.
Tests have been performed on two devices, one is the device of Layout A21-21 (5$\times$5 pixel array with the pixel size of 35$\times$35 $\mu$m),
the other one is the device of Layout A21-43 (3$\times$3 pixel array with the pixel size of 25$\times$25 $\mu$m).

\begin{figure}[htbp]
\centering
\subfloat[]{\includegraphics[width=0.5\columnwidth]{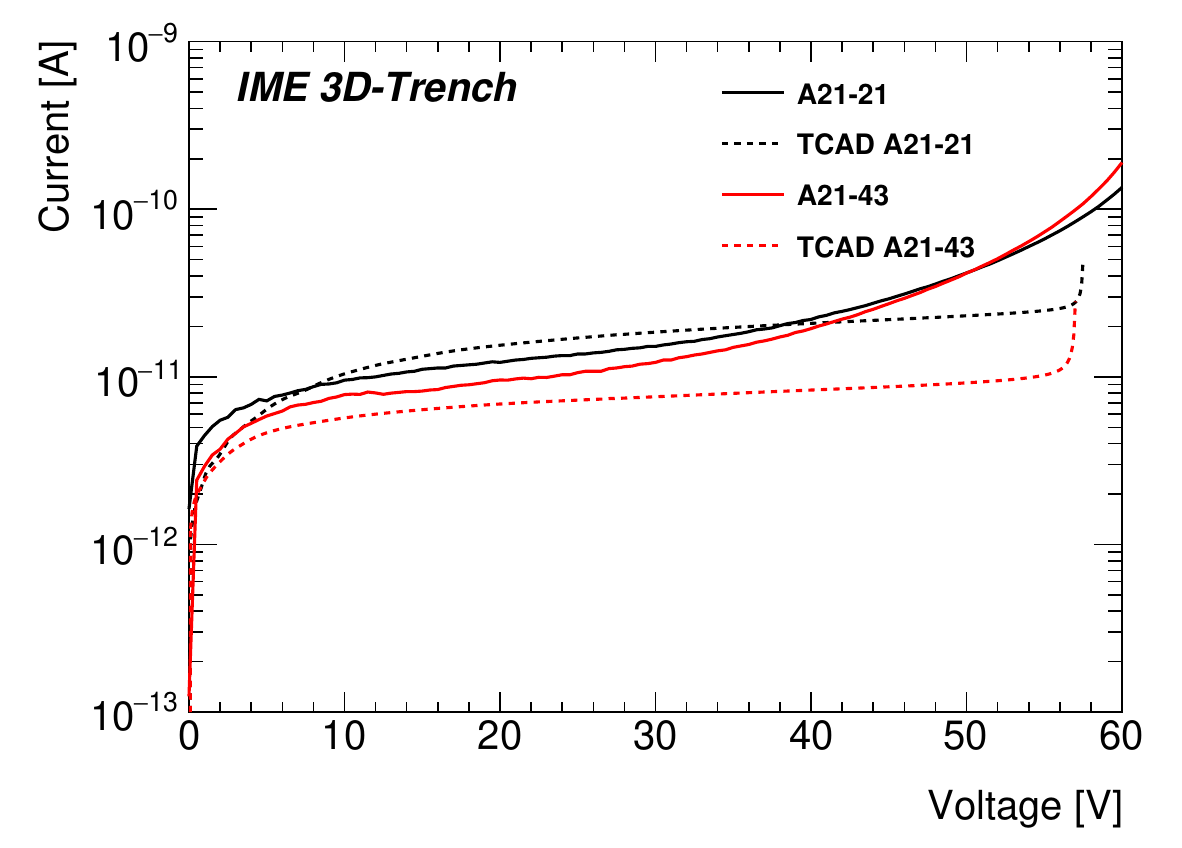} \label{fig:iv}}
\subfloat[]{\includegraphics[width=0.5\columnwidth]{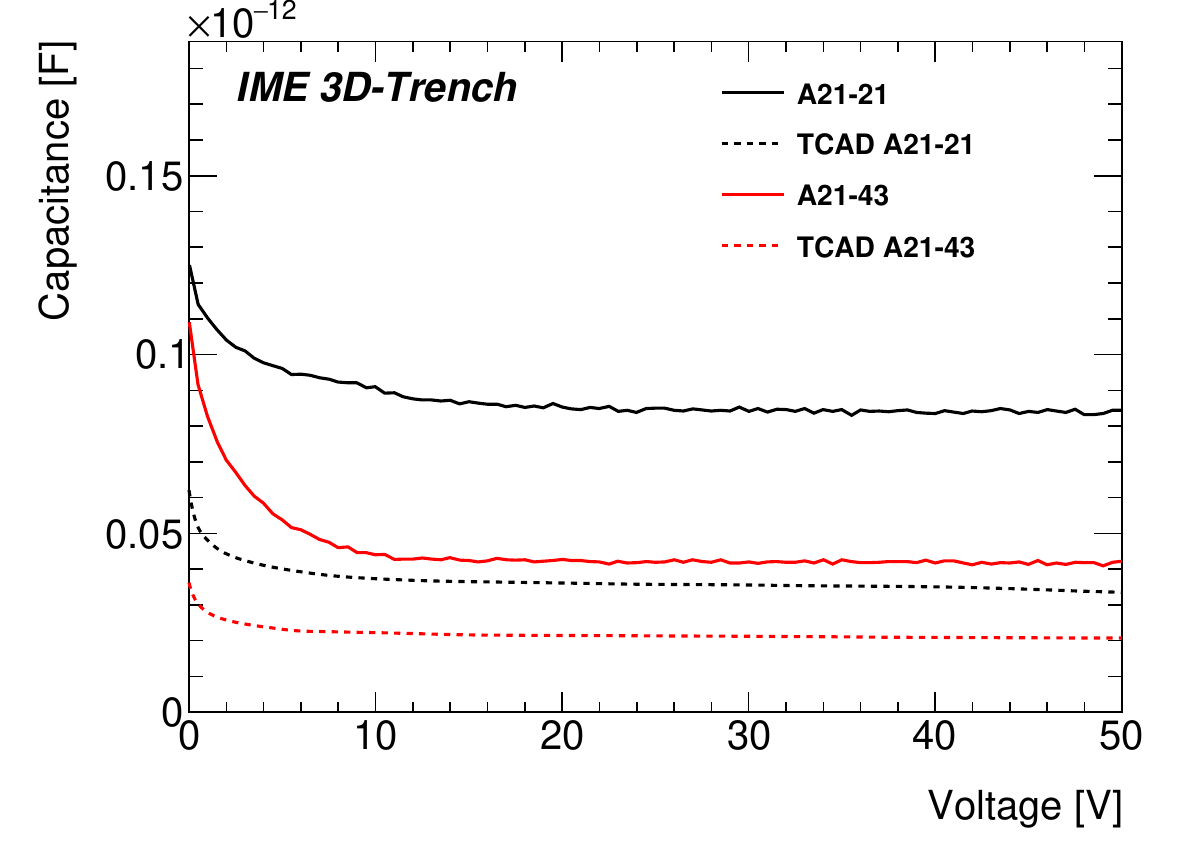} \label{fig:cv}}
\caption{The measured and simulated (a) reverse IV curves and (b) CV curves of 3D-Trench devices of the Layout A21-21 and A21-43.}
\label{fig:iv_cv}
\end{figure}

Fig. \ref{fig:iv} shows the measured and simulated reverse IV of these two devices.
The simulations agree well with the measured results up to the onset of breakdown.
%The measured breakdown voltage is $\sim$60V, whereas the simulations predict a lower breakdown voltage of $\sim$46 V, believed to be due to the difference in sharpness between the simulated and fabricated junction at the bottom of the cathode.
The reverse leakage current of the 3D-Trench device of Layout A21-21 is less than 1.22$\times$10\textsuperscript{-11} A and that of the device of Layout A21-43 is less than 9.58$\times$10\textsuperscript{-12} A up to the bias voltage of 20 V.

Fig. \ref{fig:cv} shows the measured and simulated CV of A21-21 and A21-43 pixel devices.
The CV tests were performed using an AC signal amplitude of 30 mV and 1 MHz frequency.
%The measured result agrees well with the simulation.
The discrepancy between the measurement and simulation is believed to be due to the actual cathode depth is less than 25 $\mu$m, this will be confirmed by the futuree SEM scan.
The measured saturated capacitance of these two devices are 8.4$\times$10\textsuperscript{-14} F and 4.2$\times$10\textsuperscript{-14} F, respectively. The extrapolated full depletion voltages are $\sim$10 V.

\begin{figure}[htbp]
\centering
\includegraphics[width=\columnwidth]{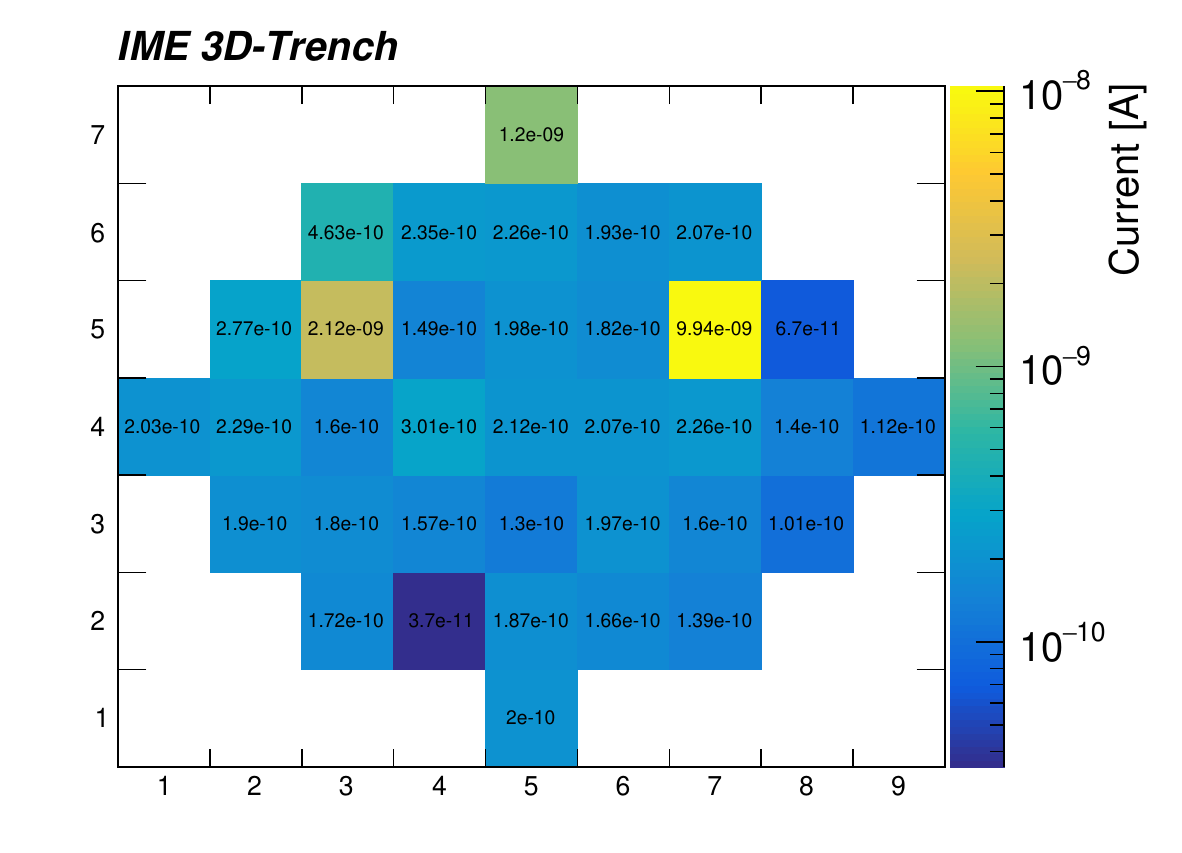}
\caption{Wafer uniformity.}
\label{fig:wafer_uniformity}
\end{figure}

Fig. \ref{fig:wafer_uniformity} shows the map of the leakage current of the A21-21 device on 8$''$ wafer before dicing. It can be seen that the largest leakage current is 9.94$\times$10\textsuperscript{-9} A, and the smallest value is 3.70$\times$10\textsuperscript{-11} A. The median is 1.93$\times$10\textsuperscript{-10} A and the mean value is 5.50$\times$10\textsuperscript{-10} A. The variance and standard deviation are 2.80$\times$10\textsuperscript{-18} A and 1.67$\times$10\textsuperscript{-9} A.
The wafer-scale leakage current measured shows good uniformity and low dispersion, with a device yield of almost 100\%.

\subsection{Infrared Laser Illumination for CCE, Time Resolution and Rise Time}
\label{subsec:laser}
\begin{figure}[htbp]
\centering
\includegraphics[width=\columnwidth]{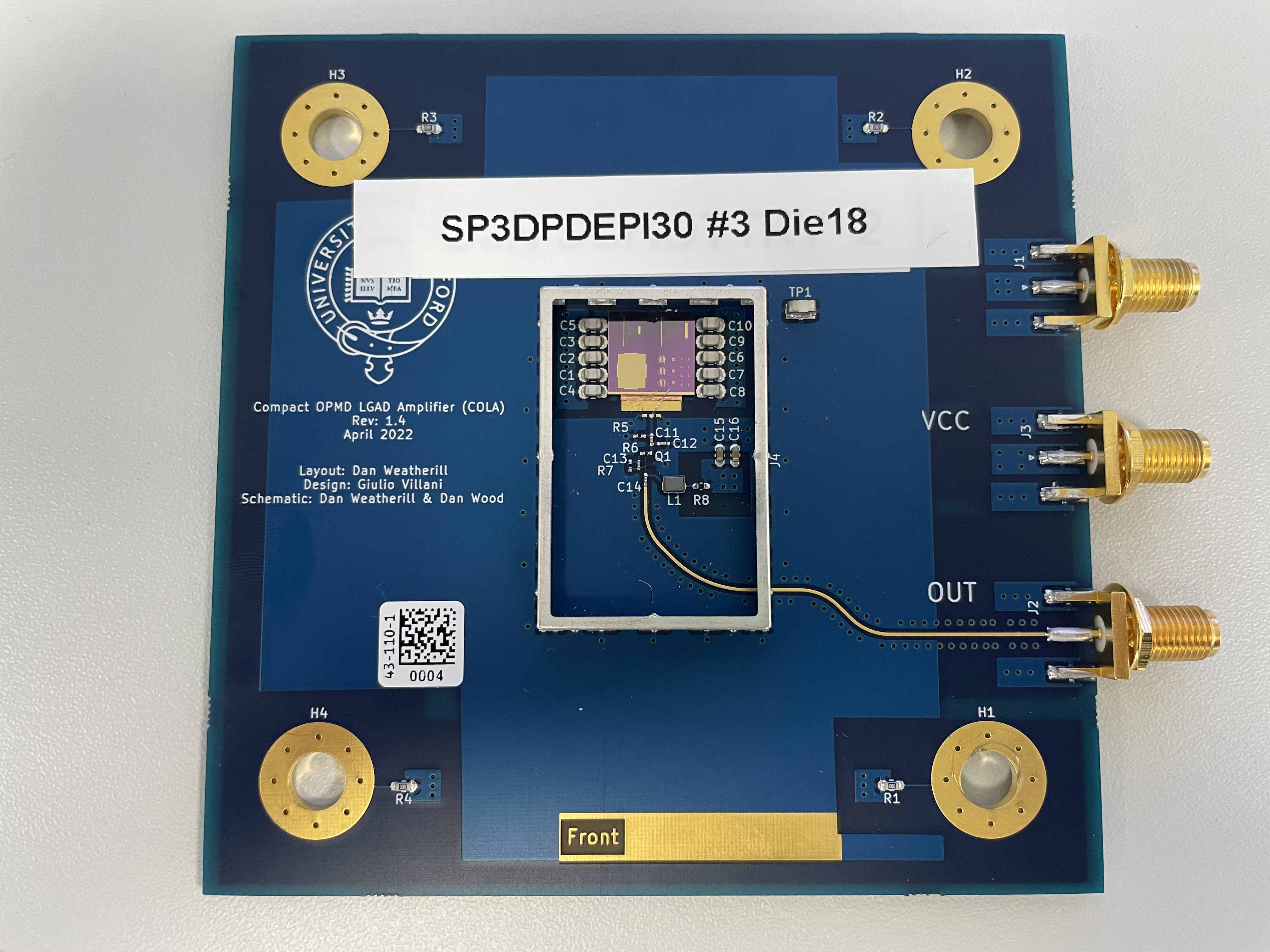}
\caption{The DUT is placed on a custom Transimpedance Amplifier (TIA) board designed at OPMD, the Compact OPMD LGAD Amplifier (COLA).}
\label{fig:cola}
\end{figure}

Measurements of the charge collection efficiency (CCE), time resolution and rise time of the 3D-Trench sensors were performed at the Rutherford Appleton Laboratory, using a customized Laser system operating in the infrared ($\lambda$ = 1064 nm), capable to inject charge with a micron spatial resolution.
The device under test (DUT) was mounted on a custom Transimpedance Amplifier COLA (Compact OPMD LGAD Amplifier, designed at the Oxford Physics Microstructure Laboratory OPMD) board, which provides a gain of 50 dB up to 1 GHz, shown in Fig. \ref{fig:cola}.
%The device under test (DUT) was mounted on a custom Transimpedance Amplifier (TIA) board designed at the Oxford Physics Microstructure Detector laboratory (OPMD) of the University of Oxford, the Compact OPMD LGAD Amplifier (COLA) with a gain of 50 dB, as shown in Fig. \ref{fig:cola}.
The sensor was biased by a Keithley 2410 HV Power Supply and signals were collected by a LeCroy WaveRunner 6100A Oscilloscope (Sampling Rate: 5 GS/s, $\sigma_{TDC}$ = 57.7 ps).

\begin{figure}[htbp]
\centering
\includegraphics[width=\columnwidth]{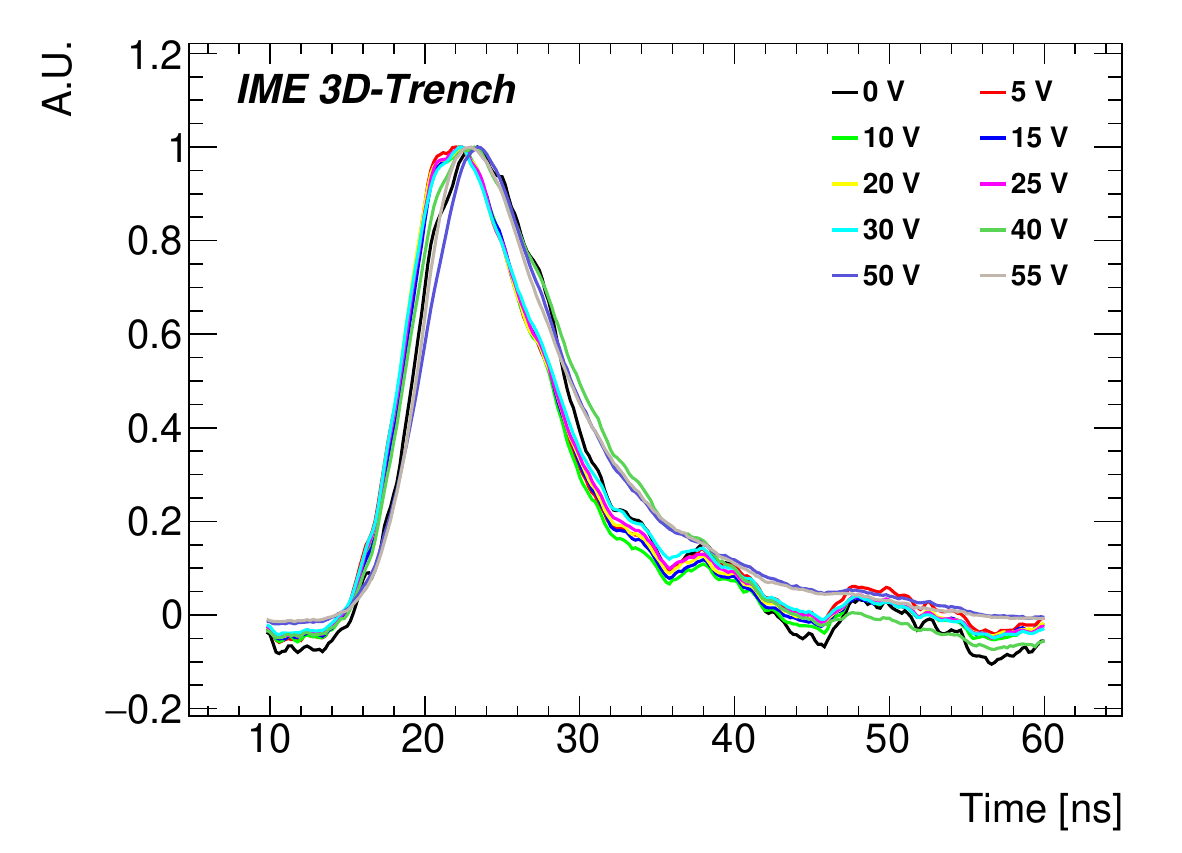}
\caption{The normalized averaged signal waveforms at various bias voltages ranging from 0 to 55 V.}
\label{fig:waveform}
\end{figure}

In this test, charge was injected using the IR laser at room temperature.
Two devices were tested, one is the device of Layout A21-21, the other one is the device of Layout A21-43.
Two different laser intensities were used for the test: Laser30 and Laser20, with the former of higher intensity and beam size big enough to illuminate the entire device array.
%Two laser densities have been used: Laser30 and Laser20. The Laser30 configuration has larger laser density meaning larger charge injection in the sensor, and the laser beam profile is large enough to illuminate the whole device surface instead of single pixel. Due to the space limitation, only 15 pixels of the 5$\times$5 pixel array were connected to the TIA, and all 9 pixels of the 3$\times$3 pixel array were connected to the TIA.
Up to 15 pixels of the 5$\times$5 pixel array and 9 of the 3$\times$3 pixel array were connected to the TIA.
Fig. \ref{fig:waveform} shows the normalized averaged signal waveforms at various bias voltages ranging from 0 to 55 V, all signal shapes agree well indicating its independence of the collected signal charge.

\begin{figure}[htbp]
\centering
\subfloat[]{\includegraphics[width=0.5\columnwidth]{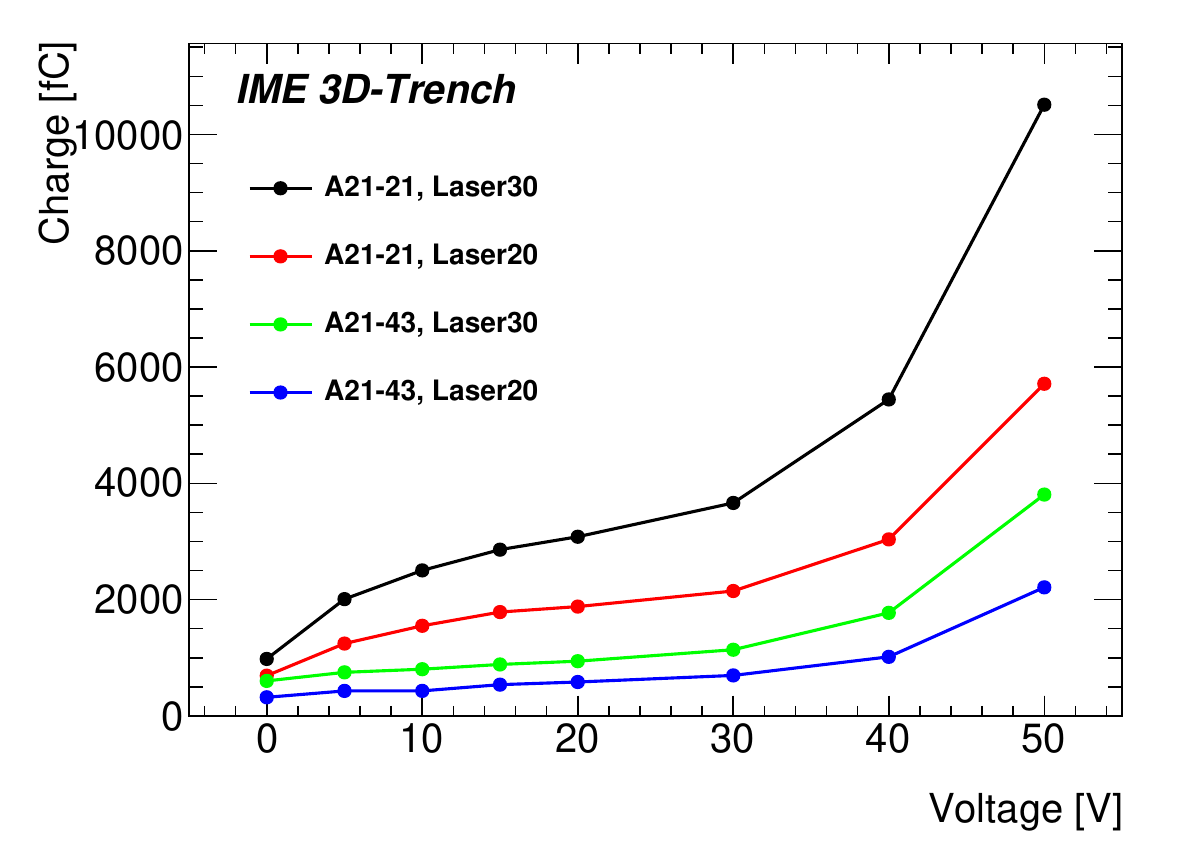}\label{fig:cce_a}}
\subfloat[]{\includegraphics[width=0.5\columnwidth]{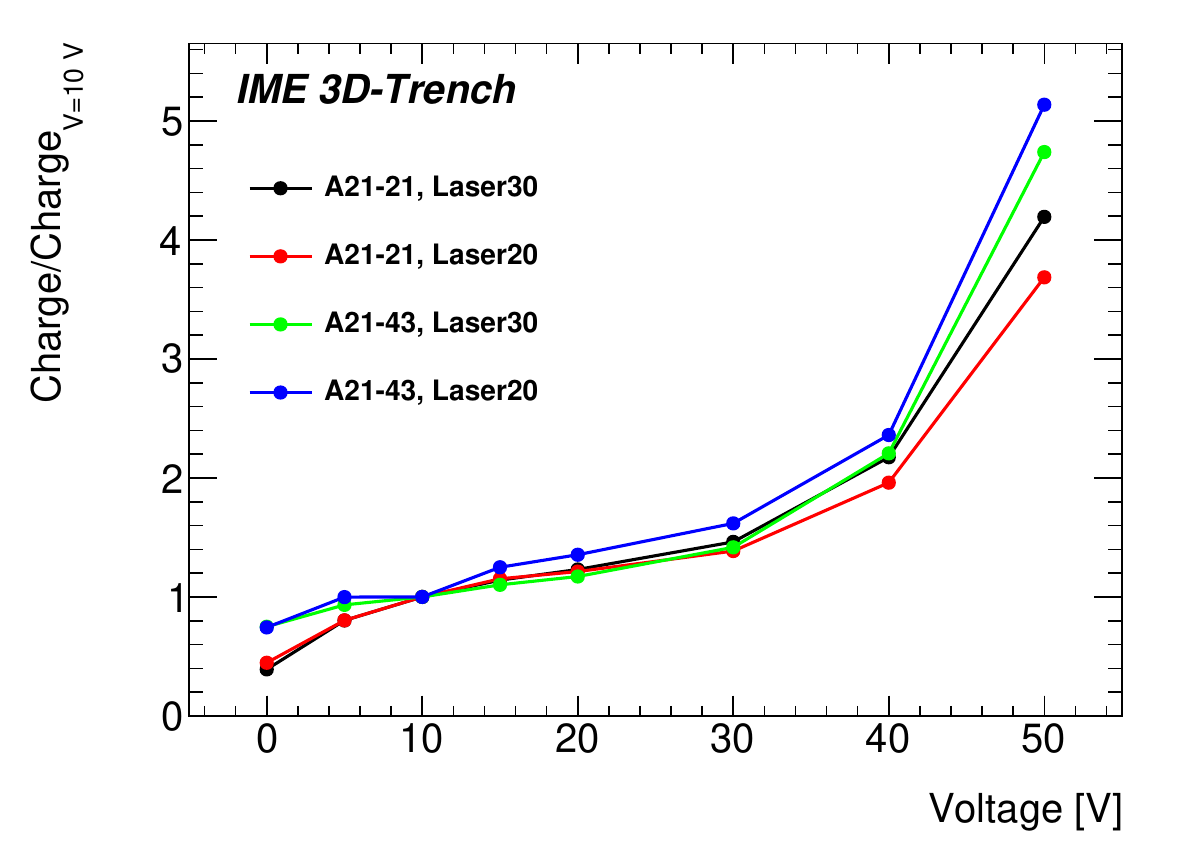}\label{fig:cce_b}}
\caption{(a) Collected charge at the bias voltage ranging from 0 to 50 V of the device of the Layout of A21-21 (5$\times$5 pixel array with the pixel size of 35$\times$35 $\mu$m) and A21-43 (3$\times$3 pixel array with the pixel size of 25$\times$25 $\mu$m), (b) Scale the collected charge to the charge collected at the bias voltage of 10 V.}
\label{fig:cce}
\end{figure}

Fig. \ref{fig:cce_a} shows the CCEs of these two 3D-Trench pixel sensors at the bias voltages ranging from 0 to 50 V. The collected charge is estimated by integration of the output pulse within a gating window and dividing the result by the nominal gain of the TIA.
Since the full depletion is achieved at $\sim$10 V, the CCE does not increase significantly from 10 to 40 V.
At the same laser intensity and bias voltage, the ratio between CCEs of the 3$\times$3 pixel sensor and the 5$\times$5 pixel sensor is $\sim$30\% which is consistent with the ratio of their areas.
Fig. \ref{fig:cce_b} shows the CCEs of these two devices scaled to their own CCE at the bias voltage of 10 V, the significant increase of the CCE from 40 V to 50 V is due to the electron impact ionization in the cathode bottom area where the electric field is high enough~\cite{Ershov_1995} to activate the electron impact ionization leading to charge multiplication.
%The steady increase from 10 V to 30 V is due to the drifting of holes in the weak electric field (the area far from the cathode), and the significant increase from 40 V to 50 V is due to the electron impact ionization in the cathode area where the electric field is high enough~\cite{Ershov_1995} to activate the electron impact ionization leading to charge multiplication.
The pixel of the 3$\times$3 pixel sensor is smaller than the pixel of the 5$\times$5 pixel sensor, leading to an early onset of impact ionisation.

\begin{figure}[htbp]
\centering
\subfloat[]{\includegraphics[width=0.5\columnwidth]{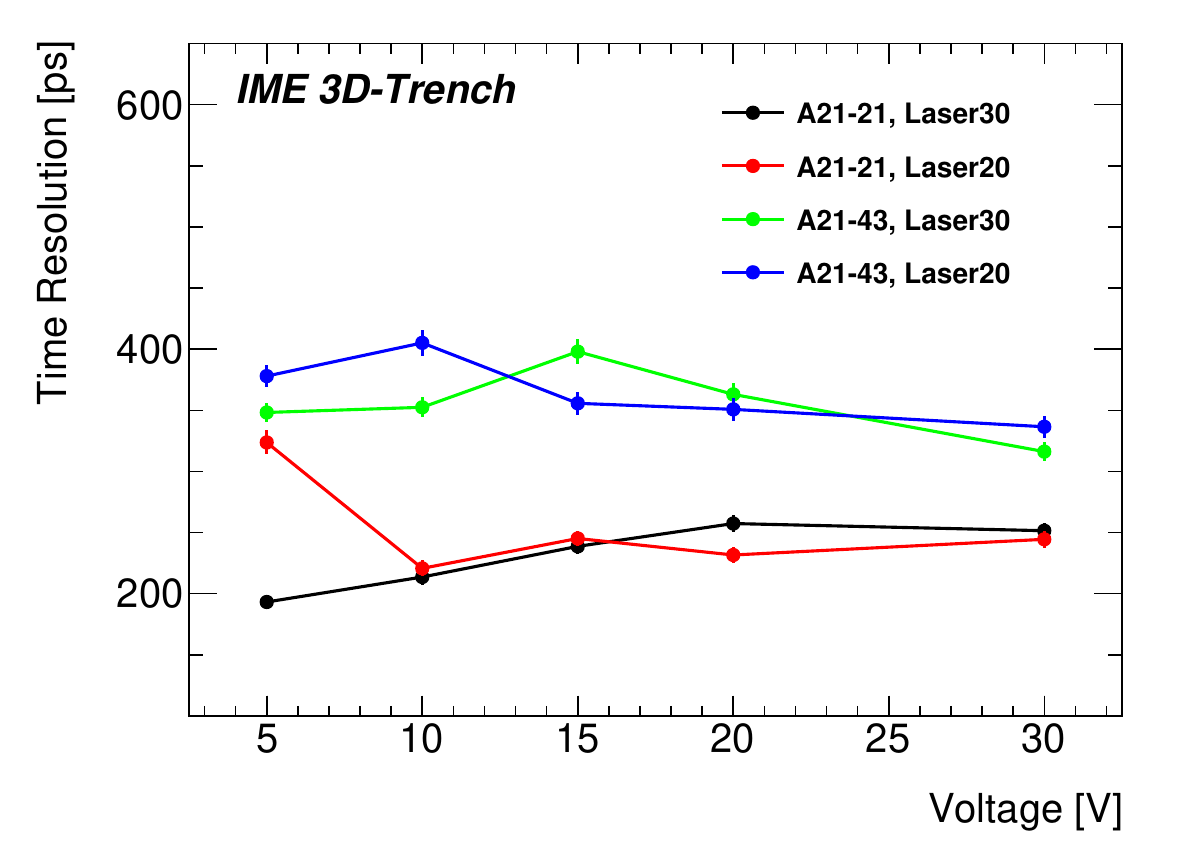}\label{fig:timeresolution}}
\subfloat[]{\includegraphics[width=0.5\columnwidth]{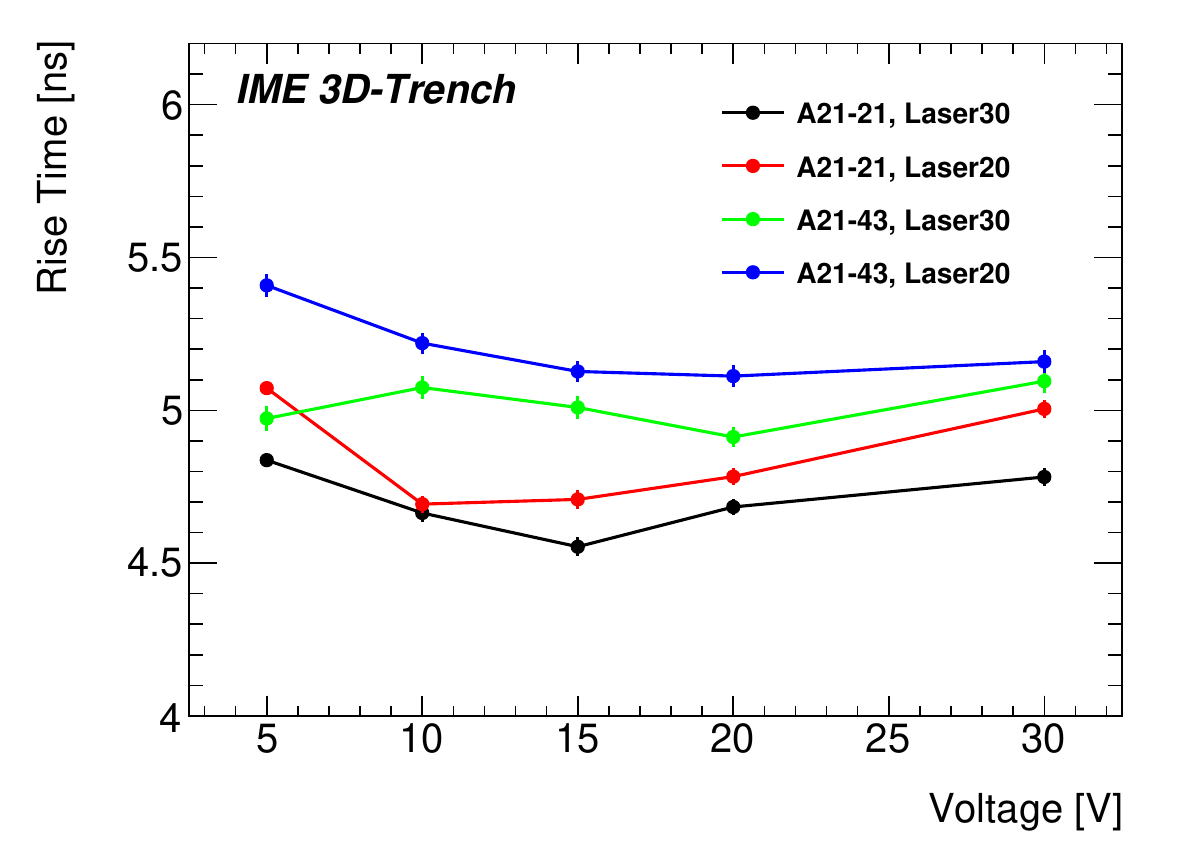}\label{fig:risetime}}
\caption{(a) Time Resolution of the device of the Layout of A21-21 (5$\times$5 pixel array with the pixel size of 35$\times$35 $\mu$m) and A21-43 (3$\times$3 pixel array with the pixel size of 25$\times$25 $\mu$m), (b) Rise Time of these two devices.}
\end{figure}

Preliminary tests were performed to measure the time resolution of the 3D-Trench sensor, from the standard deviation of the time interval between the laser trigger and a fixed threshold of the signal output. To compensate for the time walk effect, a Constant Fraction Discrimination (CFD) technique of threshold value of 30\% was applied to the signal pulse.
%The time resolution was measured using the time interval between trigger and the signal pulse. To compensate for the time walk effect from the signal pulse, a Constant Fraction
%Discriminator (CFD) technique was applied to the signal pulse. In this measurement, the CFD threshold of 30\% was used.
Fig. \ref{fig:timeresolution} shows the measured time resolutions of the 3$\times$3 pixel sensor and 5$\times$5 pixel sensor at the bias voltages ranging from 5 to 30 V at which the CCEs are relatively stable.
Both devices show a time resolution of $\sim$400 ps or better, a value believed to be comparable to the inherent timing uncertainties of the laser system. More accurate timing measurements will be performed soon.

Fig. \ref{fig:risetime} shows the measured rise time defined as the time interval from 10\% to 90\% of the pulse height at the bias voltages ranging from 5 to 30 V. Both devices show a fast rise time ($<5.5$ ns) at the current charge injection scheme, and the measured rise time shows proportional relationship with the injected charge.
Based on such fast rise time and the output of the amplifier returns to baseline within $\sim$40 ns, as shown in \ref{fig:waveform}, this kind of 3D-Trench sensor has the capacity to be used as a fast counting detector with the counting rate of $\sim$2.5 MHz or more.

%---------------
\section{Conclusion and Future Plans}
\label{sec:conclusion}

A novel 3D-Trench sensor featuring a deep trench enclosing a central column electrode has been designed and successfully fabricated. The devices have been fabricated at the 8 inch COMS pilot line at the Institute of Microelectronics of the Chinese Academy of Sciences (IMECAS).
The 3D-Trench devices were fabricated on a 30 $\mu$m thick \textit{p}-type epitaxial layer with a doping concentration of $2\times10^{13}$ cm\textsuperscript{-3} grown on the 8$''$ wafer.
Ultra-narrow etch width of 0.5 $\mu$m and high aspect ratio $>$70 have been demonstrated. Initial IV, CV, CCE and timing measurements have been carried out at IMECAS and RAL
The preliminary results shown in this paper demonstrate that the devices are successfully working. Further tests are planned to investigate their characteristics in more detail and to further validate the matching between simulation and test results.
Future devices, incorporating lessons learned from the current batch, are currently being designed.

%-----------------
\section*{Acknowledgment}
This work is supported by the National Key R\&D Program of China under Grant 2023YFF0719600, General Program of National Natural Science Foundation of China under Grant 12375188, and the Science and Technology Facilities Council (STFC) of United Kingdom under grants ST/S000747/1 and ST/W000547/1.
We are also grateful for the support with equipment and technical personnel by the IMECAS, the School of Mathematical and Physical Sciences of the University of Sheffield, and the Rutherford Appleton Laboratory.

%----------------------
%\section*{References}
\bibliographystyle{IEEEtran}
\bibliography{refernce.bib}

\begin{thebibliography}{10}
\providecommand{\url}[1]{#1}
\csname url@samestyle\endcsname
\providecommand{\newblock}{\relax}
\providecommand{\bibinfo}[2]{#2}
\providecommand{\BIBentrySTDinterwordspacing}{\spaceskip=0pt\relax}
\providecommand{\BIBentryALTinterwordstretchfactor}{4}
\providecommand{\BIBentryALTinterwordspacing}{\spaceskip=\fontdimen2\font plus
\BIBentryALTinterwordstretchfactor\fontdimen3\font minus
  \fontdimen4\font\relax}
\providecommand{\BIBforeignlanguage}[2]{{%
\expandafter\ifx\csname l@#1\endcsname\relax
\typeout{** WARNING: IEEEtran.bst: No hyphenation pattern has been}%
\typeout{** loaded for the language `#1'. Using the pattern for}%
\typeout{** the default language instead.}%
\else
\language=\csname l@#1\endcsname
\fi
#2}}
\providecommand{\BIBdecl}{\relax}
\BIBdecl

\bibitem{PARKER1997328}
\BIBentryALTinterwordspacing
S.~Parker, C.~Kenney, and J.~Segal, ``3d — a proposed new architecture for
  solid-state radiation detectors,'' \emph{Nuclear Instruments and Methods in
  Physics Research Section A: Accelerators, Spectrometers, Detectors and
  Associated Equipment}, vol. 395, no.~3, pp. 328--343, 1997, proceedings of
  the Third International Workshop on Semiconductor Pixel Detectors for
  Particles and X-rays. [Online]. Available:
  \url{https://www.sciencedirect.com/science/article/pii/S0168900297006943}
\BIBentrySTDinterwordspacing

\bibitem{785737}
C.~Kenney, S.~Parker, J.~Segal, and C.~Storment, ``Silicon detectors with 3-d
  electrode arrays: fabrication and initial test results,'' \emph{IEEE
  Transactions on Nuclear Science}, vol.~46, no.~4, pp. 1224--1236, 1999.

\bibitem{Heggelund_2022}
\BIBentryALTinterwordspacing
A.~Heggelund, S.~Huiberts, O.~Dorholt, A.~Read, O.~Rohne, H.~Sandaker,
  M.~Lauritzen, B.~Stugu, A.~Kok, O.~Koybasi, M.~Povoli, M.~Bomben, J.~Lange,
  and A.~Rummler, ``Radiation hard 3d silicon pixel sensors for use in the
  atlas detector at the hl-lhc,'' \emph{Journal of Instrumentation}, vol.~17,
  no.~08, p. P08003, aug 2022. [Online]. Available:
  \url{https://dx.doi.org/10.1088/1748-0221/17/08/P08003}
\BIBentrySTDinterwordspacing

\bibitem{Furelos_2017}
\BIBentryALTinterwordspacing
D.~V. Furelos, M.~Carulla, E.~Cavallaro, F.~Förster, S.~Grinstein, J.~Lange,
  I.~L. Paz, M.~Manna, G.~Pellegrini, D.~Quirion, and S.~Terzo, ``3d sensors
  for the hl-lhc,'' \emph{Journal of Instrumentation}, vol.~12, no.~01, p.
  C01026, jan 2017. [Online]. Available:
  \url{https://dx.doi.org/10.1088/1748-0221/12/01/C01026}
\BIBentrySTDinterwordspacing

\bibitem{Lange_2016}
\BIBentryALTinterwordspacing
J.~Lange, M.~C. Areste, E.~Cavallaro, F.~Förster, S.~Grinstein, I.~L. Paz,
  M.~Manna, G.~Pellegrini, D.~Quirion, S.~Terzo, and D.~V. Furelos, ``3d
  silicon pixel detectors for the high-luminosity lhc,'' \emph{Journal of
  Instrumentation}, vol.~11, no.~11, p. C11024, nov 2016. [Online]. Available:
  \url{https://dx.doi.org/10.1088/1748-0221/11/11/C11024}
\BIBentrySTDinterwordspacing

\bibitem{10.1063/1.4926962}
\BIBentryALTinterwordspacing
C.~Guardiola, D.~Quirion, G.~Pellegrini, C.~Fleta, S.~Esteban, M.~A.
  Cortés-Giraldo, F.~Gómez, T.~Solberg, A.~Carabe, and M.~Lozano,
  ``{Silicon-based three-dimensional microstructures for radiation dosimetry in
  hadrontherapy},'' \emph{Applied Physics Letters}, vol. 107, no.~2, p. 023505,
  07 2015. [Online]. Available: \url{https://doi.org/10.1063/1.4926962}
\BIBentrySTDinterwordspacing

\bibitem{9339896}
D.~Bachiller-Perea, J.~G. López, M.~d.~C. Jiménez-Ramos, F.~Gómez, C.~Fleta,
  D.~Quirion, A.~García-Osuna, and C.~Guardiola, ``Characterization of the
  charge collection efficiency in silicon 3-d-detectors for microdosimetry,''
  \emph{IEEE Transactions on Instrumentation and Measurement}, vol.~70, pp.
  1--11, 2021.

\bibitem{s41598-022-14940-1}
\BIBentryALTinterwordspacing
D.~Bachiller-Perea, M.~Zhang, C.~Fleta, D.~Quirion, D.~Bassignana,
  F.~G{\'o}mez, and C.~Guardiola, ``Microdosimetry performance of the first
  multi-arrays of 3d-cylindrical microdetectors,'' \emph{Scientific Reports},
  vol.~12, no.~1, p. 12240, 2022. [Online]. Available:
  \url{https://doi.org/10.1038/s41598-022-14940-1}
\BIBentrySTDinterwordspacing

\bibitem{10238431}
T.~Ariyoshi and T.~Matsunaga, ``Balanced high detection efficiency and rapid
  detection response in a silicon trench hard x-ray photon sensor,'' \emph{IEEE
  Sensors Journal}, vol.~23, no.~20, pp. 24\,465--24\,472, 2023.

\bibitem{8089762}
L.~T. Tran, L.~Chartier, D.~A. Prokopovich, D.~Bolst, M.~Povoli, A.~Summanwar,
  A.~Kok, A.~Pogossov, M.~Petasecca, S.~Guatelli, M.~I. Reinhard, M.~Lerch,
  M.~Nancarrow, N.~Matsufuji, M.~Jackson, and A.~B. Rosenfeld, ``Thin silicon
  microdosimeter utilizing 3-d mems fabrication technology: Charge collection
  study and its application in mixed radiation fields,'' \emph{IEEE
  Transactions on Nuclear Science}, vol.~65, no.~1, pp. 467--472, 2018.

\bibitem{9081916}
G.~Paternoster, G.~Borghi, M.~Boscardin, N.~Cartiglia, M.~Ferrero,
  F.~Ficorella, F.~Siviero, A.~Gola, and P.~Bellutti, ``Trench-isolated low
  gain avalanche diodes (ti-lgads),'' \emph{IEEE Electron Device Letters},
  vol.~41, no.~6, pp. 884--887, 2020.

\bibitem{10019414}
S.~Shimada, Y.~Otake, S.~Yoshida, Y.~Jibiki, M.~Fujii, S.~Endo, R.~Nakamura,
  H.~Tsugawa, Y.~Fujisaki, K.~Yokochi, J.~Iwase, K.~Takabayashi, H.~Maeda,
  K.~Sugihara, K.~Yamamoto, M.~Ono, K.~Ishibashi, S.~Matsumoto, H.~Hiyama, and
  T.~Wakano, ``A spad depth sensor robust against ambient light: The importance
  of pixel scaling and demonstration of a 2.5$\mu$m pixel with 21.8\% pde at
  940nm,'' in \emph{2022 International Electron Devices Meeting (IEDM)}, 2022,
  pp. 37.3.1--37.3.4.

\bibitem{PELLEGRINI201969}
\BIBentryALTinterwordspacing
G.~Pellegrini, M.~Manna, and D.~Quirion, ``3d-si single sided sensors for the
  innermost layer of the atlas pixel upgrade,'' \emph{Nuclear Instruments and
  Methods in Physics Research Section A: Accelerators, Spectrometers, Detectors
  and Associated Equipment}, vol. 924, pp. 69--72, 2019, 11th International
  Hiroshima Symposium on Development and Application of Semiconductor Tracking
  Detectors. [Online]. Available:
  \url{https://www.sciencedirect.com/science/article/pii/S016890021830723X}
\BIBentrySTDinterwordspacing

\bibitem{Terzo_2022}
\BIBentryALTinterwordspacing
S.~Terzo, J.~Carlotto, S.~Grinstein, M.~Manna, G.~Pellegrini, and D.~Quirion,
  ``Performance of radiation hard 3d pixel sensors for the upgrade of the atlas
  inner tracker,'' \emph{Journal of Physics: Conference Series}, vol. 2374,
  no.~1, p. 012168, nov 2022. [Online]. Available:
  \url{https://dx.doi.org/10.1088/1742-6596/2374/1/012168}
\BIBentrySTDinterwordspacing

\bibitem{DIEHL2024169517}
\BIBentryALTinterwordspacing
L.~Diehl, S.~Argyropoulos, O.~Ferrer, M.~Hauser, K.~Jakobs, M.~King, F.~Lex,
  G.~Kramberger, U.~Parzefall, G.~Pellegrini, C.~Schwemmbauer, and D.~Sperlich,
  ``Evaluation of 3d sensors for fast timing applications,'' \emph{Nuclear
  Instruments and Methods in Physics Research Section A: Accelerators,
  Spectrometers, Detectors and Associated Equipment}, vol. 1065, p. 169517,
  2024. [Online]. Available:
  \url{https://www.sciencedirect.com/science/article/pii/S0168900224004431}
\BIBentrySTDinterwordspacing

\bibitem{Dorholt_2018}
\BIBentryALTinterwordspacing
O.~Dorholt, T.~Hansen, A.~Heggelund, A.~Kok, N.~Pacifico, O.~Rohne,
  H.~Sandaker, B.~Stugu, Z.~Yang, M.~Bomben, A.~Rummler, and J.~Weingarten,
  ``Beam tests of silicon pixel 3d-sensors developed at sintef,'' \emph{Journal
  of Instrumentation}, vol.~13, no.~08, p. P08020, aug 2018. [Online].
  Available: \url{https://dx.doi.org/10.1088/1748-0221/13/08/P08020}
\BIBentrySTDinterwordspacing

\bibitem{Terzo_2021}
S.~Terzo, M.~Boscardin, J.~Carlotto, G.-F. Dalla~Betta, G.~Darbo, O.~Dorholt,
  F.~Ficorella, G.~Gariano, C.~Gemme, G.~Giannini, S.~Grinstein, A.~Heggelund,
  S.~Huiberts, A.~Kok, O.~Koybasi, A.~Lapertosa, M.~E. Lauritzen, M.~Manna,
  R.~Mendicino, H.~Oide, G.~Pellegrini, M.~Povoli, D.~Quirion, O.~M. Rohne,
  S.~Ronchin, H.~Sandaker, M.~A. Abdulla~Samy, B.~Stugu, and L.~Vannoli,
  ``Novel 3d pixel sensors for the upgrade of the atlas inner tracker,''
  \emph{Front. Phys.}, vol.~9, p. 624668, 2021.

\bibitem{6154334}
A.~Kok, M.~Boscardin, G.-F.~D. Betta, C.~Da~Via, G.~Darbo, C.~Fleta, T.-E.
  Hansen, J.~Hasi, C.~Kenney, N.~Lietaer, M.~Lozano, S.~I. Parker,
  G.~Pellegrini, and A.~Summanwar, ``Results from the first prototype of large
  3d active edge sensors,'' in \emph{2011 IEEE Nuclear Science Symposium
  Conference Record}, 2011, pp. 1319--1323.

\bibitem{chips2020006}
\BIBentryALTinterwordspacing
G.-F. Dalla~Betta and J.~Ye, ``Silicon radiation detector technologies: From
  planar to 3d,'' \emph{Chips}, vol.~2, no.~2, pp. 83--101, 2023. [Online].
  Available: \url{https://www.mdpi.com/2674-0729/2/2/6}
\BIBentrySTDinterwordspacing

\bibitem{GRENIER201133}
\BIBentryALTinterwordspacing
P.~Grenier, G.~Alimonti, M.~Barbero, R.~Bates, E.~Bolle, M.~Borri,
  M.~Boscardin, C.~Buttar, M.~Capua, M.~Cavalli-Sforza, M.~Cobal,
  A.~Cristofoli, G.-F. {Dalla Betta}, G.~Darbo, C.~{Da Vià}, E.~Devetak,
  B.~DeWilde, B.~{Di Girolamo}, D.~Dobos, K.~Einsweiler, D.~Esseni, S.~Fazio,
  C.~Fleta, J.~Freestone, C.~Gallrapp, M.~Garcia-Sciveres, G.~Gariano,
  C.~Gemme, M.-P. Giordani, H.~Gjersdal, S.~Grinstein, T.~Hansen, T.-E. Hansen,
  P.~Hansson, J.~Hasi, K.~Helle, M.~Hoeferkamp, F.~Hügging, P.~Jackson,
  K.~Jakobs, J.~Kalliopuska, M.~Karagounis, C.~Kenney, M.~Köhler, M.~Kocian,
  A.~Kok, S.~Kolya, I.~Korokolov, V.~Kostyukhin, H.~Krüger, A.~{La Rosa},
  C.~Lai, N.~Lietaer, M.~Lozano, A.~Mastroberardino, A.~Micelli, C.~Nellist,
  A.~Oja, V.~Oshea, C.~Padilla, P.~Palestri, S.~Parker, U.~Parzefall, J.~Pater,
  G.~Pellegrini, H.~Pernegger, C.~Piemonte, S.~Pospisil, M.~Povoli, S.~Roe,
  O.~Rohne, S.~Ronchin, A.~Rovani, E.~Ruscino, H.~Sandaker, S.~Seidel,
  L.~Selmi, D.~Silverstein, K.~Sjøbæk, T.~Slavicek, S.~Stapnes, B.~Stugu,
  J.~Stupak, D.~Su, G.~Susinno, R.~Thompson, J.-W. Tsung, D.~Tsybychev,
  S.~Watts, N.~Wermes, C.~Young, and N.~Zorzi, ``Test beam results of 3d
  silicon pixel sensors for the atlas upgrade,'' \emph{Nuclear Instruments and
  Methods in Physics Research Section A: Accelerators, Spectrometers, Detectors
  and Associated Equipment}, vol. 638, no.~1, pp. 33--40, 2011. [Online].
  Available:
  \url{https://www.sciencedirect.com/science/article/pii/S0168900211003524}
\BIBentrySTDinterwordspacing

\bibitem{6522814}
G.~Giacomini, A.~Bagolini, M.~Boscardin, G.-F. Dalla~Betta, F.~Mattedi,
  M.~Povoli, E.~Vianello, and N.~Zorzi, ``Development of double-sided
  full-passing-column 3d sensors at fbk,'' \emph{IEEE Transactions on Nuclear
  Science}, vol.~60, no.~3, pp. 2357--2366, 2013.

\bibitem{4696595}
A.~Zoboli, M.~Boscardin, L.~Bosisio, G.-F. Dalla~Betta, C.~Piemonte,
  S.~Ronchin, and N.~Zorzi, ``Double-sided, double-type-column 3-d detectors:
  Design, fabrication, and technology evaluation,'' \emph{IEEE Transactions on
  Nuclear Science}, vol.~55, no.~5, pp. 2775--2784, 2008.

\bibitem{5873785}
G.-F.~D. Betta, A.~Bagolini, M.~Boscardin, L.~Bosisio, P.~Gabos, G.~Giacomini,
  C.~Piemonte, M.~Povoli, E.~Vianello, and N.~Zorzi, ``Development of modified
  3d detectors at fbk,'' in \emph{IEEE Nuclear Science Symposium \& Medical
  Imaging Conference}, 2010, pp. 382--387.

\bibitem{Sultan_2017}
\BIBentryALTinterwordspacing
D.~Sultan, G.-F.~D. Betta, R.~Mendicino, M.~Boscardin, S.~Ronchin, and
  N.~Zorzi, ``First production of new thin 3d sensors for hl-lhc at fbk,''
  \emph{Journal of Instrumentation}, vol.~12, no.~01, p. C01022, jan 2017.
  [Online]. Available: \url{https://dx.doi.org/10.1088/1748-0221/12/01/C01022}
\BIBentrySTDinterwordspacing

\bibitem{LI201190}
\BIBentryALTinterwordspacing
Z.~Li, ``New bnl 3d-trench electrode si detectors for radiation hard detectors
  for slhc and for x-ray applications,'' \emph{Nuclear Instruments and Methods
  in Physics Research Section A: Accelerators, Spectrometers, Detectors and
  Associated Equipment}, vol. 658, no.~1, pp. 90--97, 2011, rESMDD 2010.
  [Online]. Available:
  \url{https://www.sciencedirect.com/science/article/pii/S0168900211008473}
\BIBentrySTDinterwordspacing

\bibitem{Diehl_2022}
\BIBentryALTinterwordspacing
L.~Diehl, O.~Ferrer, M.~Hauser, K.~Jakobs, M.~King, G.~Kramberger, N.~Moffat,
  U.~Parzefall, G.~Pellegrini, and D.~Sperlich, ``Investigation of the time
  resolution of 3d silicon sensors,'' \emph{Journal of Instrumentation},
  vol.~17, no.~12, p. C12003, dec 2022. [Online]. Available:
  \url{https://dx.doi.org/10.1088/1748-0221/17/12/C12003}
\BIBentrySTDinterwordspacing

\bibitem{TERZO2020164587}
\BIBentryALTinterwordspacing
S.~Terzo, S.~Grinstein, M.~Manna, G.~Pellegrini, and D.~Quirion, ``A new
  generation of radiation hard 3d pixel sensors for the atlas upgrade,''
  \emph{Nuclear Instruments and Methods in Physics Research Section A:
  Accelerators, Spectrometers, Detectors and Associated Equipment}, vol. 982,
  p. 164587, 2020. [Online]. Available:
  \url{https://www.sciencedirect.com/science/article/pii/S0168900220309840}
\BIBentrySTDinterwordspacing

\bibitem{DALLABETTA2016388}
\BIBentryALTinterwordspacing
G.-F. {Dalla Betta}, M.~Boscardin, M.~Bomben, M.~Brianzi, G.~Calderini,
  G.~Darbo, R.~Dell’Orso, A.~Gaudiello, G.~Giacomini, R.~Mendicino,
  M.~Meschini, A.~Messineo, S.~Ronchin, D.~Sultan, and N.~Zorzi, ``The
  infn–fbk “phase-2” r\&d program,'' \emph{Nuclear Instruments and
  Methods in Physics Research Section A: Accelerators, Spectrometers, Detectors
  and Associated Equipment}, vol. 824, pp. 388--391, 2016, frontier Detectors
  for Frontier Physics: Proceedings of the 13th Pisa Meeting on Advanced
  Detectors. [Online]. Available:
  \url{https://www.sciencedirect.com/science/article/pii/S0168900215010463}
\BIBentrySTDinterwordspacing

\bibitem{Liu_2021}
\BIBentryALTinterwordspacing
M.~Liu, S.~Lu, and Z.~Li, ``Theoretical bases of hypothetical sphere-electrode
  detectors and practical near-sphere-electrode (semisphere-electrode and
  near-semisphere-electrode) detectors,'' \emph{Journal of Physics D: Applied
  Physics}, vol.~54, no.~4, p. 045101, nov 2020. [Online]. Available:
  \url{https://dx.doi.org/10.1088/1361-6463/abbe48}
\BIBentrySTDinterwordspacing

\bibitem{Synopsys}
``Synopsys,'' \url{https://www.synopsys.com/manufacturing/tcad.html}.

\bibitem{Ershov_1995}
\BIBentryALTinterwordspacing
M.~Ershov and V.~Ryzhii, ``Temperature dependence of the electron impact
  ionization coefficient in silicon,'' \emph{Semiconductor Science and
  Technology}, vol.~10, no.~2, p. 138, feb 1995. [Online]. Available:
  \url{https://dx.doi.org/10.1088/0268-1242/10/2/003}
\BIBentrySTDinterwordspacing

\end{thebibliography}

\end{document}